\documentclass[11pt]{article}
\pdfoutput=1
\usepackage{amsmath,amssymb}
\usepackage{hyperref}
\usepackage{graphicx}
\usepackage{epstopdf}
\newcommand{\bea}{\begin{eqnarray}}
\newcommand{\eea}{\end{eqnarray}}

\newcommand\CZ{{\cal{Z}}}

\newsavebox{\uuunit}
\sbox{\uuunit}
    {\setlength{\unitlength}{0.825em}
     \begin{picture}(0.6,0.7)
        \thinlines
        \put(0,0){\line(1,0){0.5}}
        \put(0.15,0){\line(0,1){0.7}}
        \put(0.35,0){\line(0,1){0.8}}
       \multiput(0.3,0.8)(-0.04,-0.02){10}{\rule{0.5pt}{0.5pt}}
     \end {picture}}

\numberwithin{equation}{section}

\begin{document}

\thispagestyle{empty}

\hfill LMU-ASC 76/12\\[-3ex]

\hfill MPP-2012-145 \\
\vskip 6mm

\begin{center}
{\bf\LARGE
Extremal black brane solutions in  \\ \vskip 2mm
five-dimensional gauged supergravity}

\vspace{10mm}

{\large
{\bf Susanne Barisch-Dick$^{*,\times}$,}
{\bf Gabriel Lopes Cardoso$^+$,}
{\bf Michael Haack$^*$,}\\ \vskip 3mm
{\bf Suresh Nampuri$^\dagger$}
\vspace{1cm}

{\it $^*$
 Arnold Sommerfeld Center for Theoretical Physics \\ [1mm]
Ludwig-Maximilians-Universit\"at M\"unchen \\ [1mm] 
Theresienstrasse 37, 80333 M\"unchen, Germany \\ [5mm] 
\it $^\times$
Max-Planck-Institut f\"ur Physik \\ [1mm]
F\"ohringer Ring 6, 80805 M\"unchen, Germany \\ [5mm] 
$^+$
CAMGSD, Departamento de Matem\'atica\\ [1mm]
Instituto Superior T\'ecnico, Universidade T\'ecnica de Lisboa\\ [1mm]
Av. Rovisco Pais, 1049-001 Lisboa, Portugal \\ [5mm] 
$^\dagger$ 
Laboratoire de Physique Th\'eorique \\ [1mm]
Unit\'e Mixte du CNRS et de l'\'Ecole Normale Sup\'erieure \\[1mm]
\'Ecole Normale Sup\'erieure\\[1mm]
24 rue Lhomond, 75231 Paris Cedex 05, France 
}}

\end{center}
\vspace{8mm}

\begin{center}
{\bf ABSTRACT}\\
\end{center}

We study stationary black brane solutions in the context of $N = 2$, $U(1)$
gauged supergravity in five dimensions. Using the formalism of
first-order flow equations, we construct examples of extremal black brane
solutions that include Nernst branes, i.e.\ extremal black
brane solutions with vanishing entropy density, as well as black branes with cylindrical horizon topology,
 whose entropy density can be computed from a Cardy formula of the dual CFT.
\clearpage
\setcounter{page}{1}
\tableofcontents



\section{Introduction}

Extremal black solutions in low energy effective theories arising from string theories always offer scope for understanding aspects of the Hilbert space of the quantum gravity theory that arises in this context. In the fortuitous cases where the asymptotics of the geometry or the near horizon geometry is either global AdS or a quotient of the Poincar\'e patch of AdS, these solutions may be analyzed in terms of thermal ensembles in the holographic dual CFTs, and hence either offer a bulk view of strongly coupled field theory processes in the boundary theory or a microscopic understanding of the thermodynamic properties of the extremal black solutions \cite{Maldacena:1997re}.

There has been extensive progress in constructing and analyzing extremal black hole solutions from both the bulk and the holographic CFT points of view. Recent developments in the construction of extremal black solutions with non-trivial scalar fields in gauged supergravity theories in four dimensions show that the presence of the fluxes can give rise to a wide variety of asymptotically non-flat backgrounds \cite{Cacciatori:2009iz,Dall'Agata:2010gj,Hristov:2010ri,Kachru:2011ps,
Barisch:2011ui,Donos:2011pn,Meessen:2012sr}. 
One of the salient features of the solution space of gauged supergravity actions in four dimensions is the existence of 
horizons with non-spherical horizon topology, such as
$\mathbb{R}^2$, and a specific subset of these solutions involve extremal black branes with zero area density and hence zero entropy density \cite{Goldstein:2009cv,Goldstein:2010aw,Berglund:2011cp,Barisch:2011ui}.
The thermodynamic behavior of these systems are closest to real condensed matter systems (low entropy at zero temperature) and in cases where these brane solutions can be found in asymptotic AdS backgrounds, they can in principle be used to study dual condensed matter systems with quantum phase transitions at zero temperature, as in \cite{D'Hoker:2010ij, Faulkner:2010gj, Erdmenger:2011hp, D'Hoker:2012ih}.
An example of an extremal black brane solution that satisfies the third law of thermodynamics (the Nernst law) was obtained in
\cite{Barisch:2011ui} as a solution to the STU model of $N = 2$, $U(1)$ 
gauged supergravity in four dimensions. However, it was also found to be 
difficult to obtain analytic solutions describing extremal black 
brane solutions in asymptotic $AdS_4$ backgrounds which, as recalled above,
represents a worthy endeavor in view of possible applications in holography.
Hence, in the following,
we shift focus to gauged supergravity in five dimensions,
with the intent of finding extremal solutions in 
asymptotic 
$AdS_5$ backgrounds, or extremal solutions with a near horizon geometry given in terms of a quotient of the Poincar\'e patch of $AdS_3$, so that one can use the dual CFT to arrive at a microscopic understanding of the black brane entropy density. 
We will not rely on supersymmetry to construct these extremal solutions.  Various types of extremal (not necessarily
supersymmetric) five-dimensional black solutions with flat horizons have already been discussed in 
\cite{
Chamblin:1999tk,Gutowski:2004ez,Kunduri:2007qy,
Grover:2008ih,D'Hoker:2009mm,D'Hoker:2009bc,D'Hoker:2010ij,Almuhairi:2010rb,Almuhairi:2011ws,Donos:2011ff,
Donos:2011pn, D'Hoker:2012ih}.

In this paper, we follow roughly the pattern of exploration set up in \cite{Dall'Agata:2010gj,Barisch:2011ui} for 
the four-dimensional case. We begin by rewriting the five-dimensional
$N = 2$, $U(1)$ gauged supergravity action in terms of squares of first-order flow equations.  In the ungauged case,
it is known \cite{Bellucci:2010aq} that there exist multiple rewritings based on different 'superpotentials',
depending on the charges that are turned on.  In the presence of fluxes, we observe a similar feature.
The flow equations we obtain are supported by 
electric charges, magnetic fields and fluxes of electric type. 
The solutions we construct include Nernst solutions in asymptotic $AdS_5$ backgrounds (i.e. black solutions with vanishing entropy density)
as well as non-Nernst black brane solutions that describe extremal BTZ$\times \mathbb{R}^2$-solutions.
The latter have a cylindrical 
horizon topology $S^1\times \mathbb{R}^2$, with the geometry being a quotient of the Poincar\'e patch of $AdS_3$ trivially fibered over an $\mathbb{R}^2$ surface. The near-horizon $AdS_3 \times \mathbb{R}^2$ solution
has been obtained before in \cite{Kunduri:2007qy,Grover:2008ih} using an analysis based on supersymmetry.

We can immediately compute the entropy density of the 
BTZ$\times \mathbb{R}^2$
black brane by using the Cardy formula of the dual CFT, thereby obtaining a microscopic derivation of the bulk entropy density. A salient aspect of the first-order rewriting that gives rise to these black branes is the fact that the angular momentum, the electric quantum numbers and the magnetic fields are organized into 
quantities which are invariant under the spectral flow of the theory, exactly as in the ungauged case \cite{deBoer:2006vg}. This serves as a useful tool to identify the real quantum numbers of a worldvolume CFT in a string theory construction of gauged
supergravity, and sets an indicator of the symmetries such a purported theory should satisfy.

We also reproduce the non-extremal black brane solutions of \cite{Behrndt:1998jd}
and the electric solutions obtained recently in \cite{Donos:2011ff,Donos:2011pn}.

The paper is organized as follows. 
We consider two first-order rewritings of the bosonic action of $N = 2$, $U(1)$ gauged supergravity.
The first rewriting is performed in section \ref{sec:first-ord}.  The solution space of the resulting
first-order flow equations encompasses static, purely magnetic solutions.  We verify that the Hamiltonian
constraint is satisfied (appendix \ref{sec:ricci}
summarizes the Einstein equations of motion). In section \ref{sec:4d5d} we briefly discuss the relation of
these flow equations with the four-dimensional flow equations obtained in \cite{Dall'Agata:2010gj,Barisch:2011ui}.
We refer to appendix \ref{45dict} for the details of the comparison.  In section \ref{sec:ex-sol}
we turn to the construction of solutions to the first-order flow equations in five dimensions.
First we consider exact solutions with constant scalar fields.  These solutions do not carry electric fields, but may
have magnetic fields, and they have rotation.  We construct extremal BTZ$\times \mathbb{R}^2$ solutions
that are supported by magnetic fields, as well as rotating Nernst geometries in asymptotic $AdS_5$ backgrounds.
Then we obtain numerical solutions with non-vanishing scalar fields, with and without rotation. These have BTZ$\times \mathbb{R}^2$
near horizon geometry and are asymptotically $AdS_5$. They constitute generalizations of  
a solution given in \cite{Donos:2011pn} to the case with several running scalar fields and rotation. 

In appendix \ref{electric-rew} we turn to a different first-order rewriting.  This is motivated by the search
for solutions with electric fields. This rewriting is the one performed in \cite{Cardoso:2008gm} for static black hole solutions, which we adapt
to the case of stationary black branes in the presence of 
magnetic fields.  The resulting first-order flow equations allow for the non-extremal black brane solutions
constructed in \cite{Behrndt:1998jd}, as well as for the extremal 
electric solutions obtained in \cite{Donos:2011ff,Donos:2011pn}.


\section{First-order flow equations for stationary solutions \label{sec:first-ord}}

In the following, we derive first-order flow equations for 
extremal stationary black brane solutions in $N = 2, U(1)$
gauged supergravity in five dimensions with $n$ Abelian vector multiplets. We work in big moduli space.
We follow the exposition given in \cite{Bellucci:2010aq} for the ungauged case and
adapt it to the gauged case.

\subsection{Flow equations in big moduli space}

Following \cite{D'Hoker:2009bc}, 
we make the ansatz for the black brane line element, 
\begin{equation}\label{metricrotation}
 ds^2=-e^{2U(r)} dt^2+e^{2V(r)}dr^2+e^{2B(r)}(dx^2+dy^2)+e^{2W(r)}(dz+C(r)dt)^2\,,
\end{equation}
while for the Abelian gauge fields $A_M^A$ ($A = 1 , \dots, n$) we take
\begin{equation}\label{gaugefielddyonic}
 A^A_M dx^M=A^A_t \,dt+P^A \, x\, dy+A^A_z\, dz=(e^A+A^A_z \, C(r) ) \,dt+P^A \, x \,dy+A^A_z\, dz\,.
\end{equation}
Here the $P^A$ are constants and $A^A_t$, $A^A_z$ depend only on $r$.
The associated
field strength components read
\begin{eqnarray}\label{gaugefield}
 F^A_{rt}&=&(A^A_t)'=(e^A)'+A^A_z \, C'+(A^A_z)' \, C \;,\nonumber\\
F^A_{xy}&=&P^A \;,\nonumber\\
F^A_{rz}&=&(A^A_z)'\,,
\end{eqnarray}
where $'$ denotes differentiation with respect to $r$, and $(e^A)'$ corresponds to the four-dimensional electric field upon dimensional reduction.  The solutions we seek will be supported by real scalar fields $X^A(r)$ and by electric
fluxes $h_A$. The ansatz \eqref{metricrotation} and \eqref{gaugefielddyonic} is the most general ansatz with translational invariance  in the coordinates $t,x,y$ and $z$ and with rotational invariance in the $x,y$-plane, cf.\ \cite{D'Hoker:2012ih}.

The bosonic part of the five-dimensional action describing $N = 2, U(1)$
gauged supergravity is given by \cite{Gunaydin:1984pf,Gunaydin:1984ak}
\begin{eqnarray}
 S&=&\int dx^5\,\bigg[\sqrt{-g}\left(R-G_{AB} \, \partial_M
 X^A \partial^M X^B-\frac12 G_{AB} F^A_{MN} F^{B\,MN} \right. \nonumber\\
&& \left. \qquad \qquad \qquad  -g^2 (G^{AB}h_Ah_B-2(h_AX^A)^2) \, \right)\nonumber\\
&&\phantom{\int dx^5\,\bigg(}-\frac{1}{24} C_{ABC}F^A_{KL}F^B_{MN}A^C_P \, \epsilon^{KLMNP}\bigg]\,,
\label{5d-action-bos}
\end{eqnarray}
where the scalar fields $X^A$ satisfy the constraint $\frac16 C_{ABC}X^AX^BX^C=1$.
The target space metric $G_{AB}$ is given by
\begin{equation}
G_{AB} = - \frac12 C_{ABC} X^C + \frac92 X_A X_B\ ,
\label{Gab-met}
\end{equation}
where 
\bea
X_A = \frac23 G_{AB} \, X^B =  \frac16 C_{ABC} X^BX^C\ .
\eea
Inserting the solution ansatz into this action, 
we find that the Ricci scalar contributes
\begin{eqnarray}
 \sqrt{-g} \, R &=& e^{2B+W+U-V}\left(2 B'^2+2U'W'+4B'W'+4B'U'+\frac12 e^{2W-2U} C'^2\right) \nonumber\\
&& -\left[2e^{2B+W+U-V}(2B'+W'+U')\right]'\,,
\end{eqnarray}
while the gauge field kinetic terms contribute
\begin{eqnarray}
&& \sqrt{-g}\left(-\frac12G_{AB}F^{A}_{MN} F^{B\,MN}\right)
=e^{2B+W+U-V}\bigg(-G_{AB}P^AP^Be^{2V-4B}+G_{AB}F^{A}_{rt}F^B_{rt}e^{-2U} \nonumber\\
&& \qquad \qquad \qquad -G_{AB}(A^A_z)'(A^B_z)'(e^{-2W}-e^{-2U}C^2)-2G_{AB}F^A_{rt}(A^B_z)'e^{-2U}C \bigg)\,,
\end{eqnarray}
with $F^A_{rt}$ given in \eqref{gaugefield}.  The Chern-Simons term, on the other hand,  can be rewritten as 
\begin{eqnarray}\label{CS-term}
 \int dx^5\left( -\frac{1}{24} C_{ABC}F^A_{KL}F^B_{MN}A^C_P \, \epsilon^{KLMNP}\right)
=\int dx^5 \left(-C_{ABC}F^A_{rt}F^B_{xy}A^C_z+TD\right)\,,
\end{eqnarray}
where $TD$ denotes a total derivative term.
Inserting these expressions into \eqref{5d-action-bos}
yields the one-dimensional Lagrangian ${\cal L}$, 
\begin{eqnarray}
{\cal L}&=&e^{2B+W+U-V}\bigg(2 B'^2+2U'W'+4B'W'+4B'U'+\frac12 e^{2W-2U} C'^2-G_{AB}(X^A)'(X^B)'\nonumber\\
&&\phantom{e^{2B+w+U-V}\bigg(}-G_{AB}P^AP^Be^{2V-4B}+G_{AB}(e^A)'(e^B)'e^{-2U}+G_{AB}A^A_zA^B_zC'^2e^{-2U}\nonumber\\
&&\phantom{e^{2B+w+U-V}\bigg(}+2G_{AB}(e^A)'A^B_z C'e^{-2U}-G_{AB}(A^A_z)'(A^B_z)'e^{-2W}\nonumber\\
&&\phantom{e^{2B+w+U-V}\bigg(}-g^2 e^{2V}(G^{AB}h_Ah_B-2(h_AX^A)^2)\bigg) \nonumber\\
&&-\,C_{ABC}(e^A)'P^BA^C_z-\,C_{ABC}A^A_zP^BA^C_z
C'-\,C_{ABC}(A^A_z)'P^BA^C_zC \,,
\label{1d-lag}
\end{eqnarray}
where we dropped total derivative terms.

Now we express the electric field $(e^A)'$ in terms of electric charges $q_A$ by performing the 
Legendre transformation ${\cal L}_L={\cal L}-q_A(e^A)'$, and obtain
\begin{equation}
 (e^A)'=\frac12 e^{-2B-W+U+V}G^{AB}\, \hat{q}_B -A^A_z C' \;,
 \label{flow-e'}
\end{equation}
where
\begin{equation} \label{hatq}
\hat{q}_A=q_A+C_{ABC}P^BA^C_z\,.
\end{equation}
Substituting this relation in \eqref{1d-lag} gives
\begin{eqnarray}\label{lagrange1}
{\cal L}&=&e^{2B+W+U-V}\bigg(2 B'^2+2U'W'+4B'W'+4B'U'+\frac12 e^{2W-2U} C'^2-G_{AB}(X^A)'(X^B)'\nonumber\\
&&\phantom{e^{2B+w+U-V}\bigg(}-G_{AB}P^AP^Be^{2V-4B}-\frac14 G^{AB}\hat{q}_A\hat{q}_B \, e^{-4B-2W+2V}\nonumber\\
&&\phantom{e^{2B+w+U-V}\bigg(}-G_{AB}(A^A_z)'(A^B_z)'e^{-2W}-g^2 e^{2V}(G^{AB}h_Ah_B-2(h_AX^A)^2)\bigg)\nonumber\\
&&-\,C_{ABC}(A^A_z)'P^BA^C_zC + q_AA^A_z C'\,.
\end{eqnarray}
Furthermore, using
\begin{eqnarray}
 (C_{ABC} A^A_z P^B A_z^C C)' = 
2 C_{ABC} (A_z^A)'P^B A_z^C C + C_{ABC}  A^A_z P^B  A^C_z C' \;,
\end{eqnarray}
we obtain 
\begin{equation}
-C_{ABC}(A^A_z)'P^BA^C_zC=\frac12 C_{ABC} A^A_zP^B A^C_z C'+TD\,,
\end{equation}
where $TD$ denotes again a total derivative, which we drop in the following.

Next, we express $C'$ in terms of a constant quantity $J$ which, in the compact
case, corresponds to angular momentum. We do this by performing the Legendre transformation
${\cal L}_L={\cal L}-JC'$,  and obtain 
\begin{equation} \label{Cprime}
 C'=e^{-2B-3W+U+V}\, \hat{J} \;,
\end{equation}
where
\begin{equation} \label{Jhat}
\hat{J}=J-q_A A^A_z-\frac12 C_{ABC} A^A_z A^B_z P^C \,.
\end{equation}
This results in 
\begin{eqnarray}\label{lagrange2}
{\cal L}&=&e^{2B+W+U-V}\bigg(2 B'^2+2U'W'+4B'W'+4B'U'-G_{AB}(X^A)'(X^B)' \nonumber\\
&& -G_{AB}(A^A_z)'(A^B_z)'e^{-2W}
-G_{AB}P^AP^Be^{2V-4B}-\frac14 G^{AB}\hat{q}_A\hat{q}_B
\, e^{-4B-2W+2V} \nonumber\\
&& -\frac12 e^{-4B-4W+2V}\hat{J}^2
-g^2 e^{2V}(G^{AB}h_Ah_B-2(h_AX^A)^2)\bigg) \,.
\end{eqnarray}

Now we rewrite the one-dimensional Lagrangian \eqref{lagrange2} as a sum of squares of first-order flow
equations.  To this end, we use the relation
\begin{eqnarray}
 \Big(e^{U-W}(\hat{q}_A A^A_z-\frac12 C_{ABC}A^A_zP^BA^C_z)\Big)'
 =e^{U-W}\Big((U'-W')(-\hat{J})+\hat{q}_A(A^A_z)'\Big)+\left(J e^{U-W}\right)'\,, \nonumber\\
\end{eqnarray}
and obtain
\begin{eqnarray}\label{dyonicrew}
{\cal L} 
&=&e^{2B+W + U-V}\bigg[-e^{-2 W} G_{AB}\Big((A^A_z)'+\frac12 G^{AC}\hat{q}_C e^{-2B+V}\Big)\Big((A^B_z)'+\frac12 G^{BD}\hat{q}_D e^{-2B+V}\Big)\nonumber\\
&&\phantom{e^{2B+W + U-V}\bigg[}-\frac12 \left(\hat{J}e^{-2B-2W+V}-(U'-W')\right)^2\nonumber\\
&&\phantom{e^{2B+W + U-V}\bigg[}
-  \left(B' - \frac12(U' + W') + \tfrac32 X_A P^A \, e^{V-2B} \right)^2 \nonumber\\
&&\phantom{e^{2B+W + U-V}\bigg[}
 + \frac13 \left( 3 \left(B' + \tfrac12(U' + W') \right) -  2g X^A h_A e^V+ \frac32 X_A \, P^A \,
 e^{V-2 B} \right)^2 \nonumber\\
&& - G_{AB} \left( X'^A - e^V \left[ \frac23 X^C (gh_C + G_{CD} P^D e^{-2B}) X^A - G^{AC} 
(gh_C +  G_{CD} P^D e^{-2B})
 \right] \right) \nonumber\\
 &&\qquad \quad \left( X'^B -  e^V \left[ \frac23 X^E (gh_E + G_{EF} P^F e^{-2B}) X^B - G^{BE} 
(gh_E +  G_{EF} P^F e^{-2B})
 \right] \right)
\bigg]
\nonumber\\
&& + 2 \left( e^{2B+W + U} \,\left(  g X^A h_A - \frac32 X_A P^A \, e^{-2B} \right)
\right)' \nonumber\\
&&
+\left(e^{U-W} (\hat{q}_A A^A_z-\frac12 C_{ABC}A^A_zP^BA^C_z)\right)'- \left(J e^{U-W}\right)'\nonumber\\
&&
+ 2ge^{W+ U+V} \, h_A P^A\,.
\end{eqnarray}
This concludes the rewriting of the effective one-dimensional Lagrangian.

Setting the squares in \eqref{dyonicrew} to zero yields the first-order flow equations
\begin{eqnarray}\label{eqndyonic}
(A^A_z)'&=&-\frac12 G^{AC}\hat{q}_C \, e^{V-2B}\,,\nonumber\\
(X^A)'&=& \frac23 X^C (gh_Ce^V + G_{CD} P^D e^{V-2B}) X^A - G^{AC} 
(gh_C e^V+  G_{CD} P^D e^{V-2B})\,,\nonumber\\
U'-W'&=&\hat{J} \, e^{-2B-2W+V}\,,\nonumber\\
0&=&B' - \frac12(U' + W') + \frac32 X_A P^A \, e^{V-2B}\,,\nonumber\\
0&=&3 \left(B' + \frac12(U' + W') \right) -  2g X^A h_A e^V+ \frac32 X_A \, P^A \,
 e^{V-2 B} \;.
\end{eqnarray}
These flow equations are supplemented by \eqref{flow-e'} and \eqref{Cprime}, and solutions to 
these equations are subjected to the constraint
\begin{equation} \label{hpconstraint}
h_AP^A=0 \;,
\end{equation}
which follows from the last line of \eqref{dyonicrew}. Note that the flow equations \eqref{eqndyonic} show an interesting decoupling: The scalar fields $X^A$ and the metric coefficient $e^B$ are completely determined by the magnetic fields and the fluxes, whereas the electric charges only enter in the equations for the metric functions $e^U$ and $e^W$ and the $A^A_z$-components of the gauge fields. This will be helpful in the search for solutions, cf.\ sec.\ \ref{sec:ex-sol}.

Subtracting the fourth from the fifth equation in \eqref{eqndyonic} gives
\begin{equation}\label{eqUBW}
B'+U'+W'=g \, h_AX^A \, e^V.
\end{equation}
When $B$ is constant, this yields a flow equation for $U +W$ that, when compared with the fourth equation of
\eqref{eqndyonic}, yields the condition
\begin{equation}
g \, X^A h_A = 3 X_A P^A \, e^{-2B} \;.
\label{B-cond}
\end{equation}

Also observe that \eqref{gaugefield}, \eqref{flow-e'} and the first equation of \eqref{eqndyonic} implies 
\begin{equation}
F_{rt}^A = \frac12 (1-\hat{C}) \,  e^{- 2 B - W + U + V} G^{AB} \hat{q}_B\;,
\label{eq:F_rt-sol}
\end{equation}
where 
\begin{equation}
{\hat C} \equiv C \, e^{-(U- W)} \;,
\label{eq:integr-const-C}
\end{equation}
while the third equation, together with \eqref{Cprime}, gives
\begin{equation}
C' = \left(U' - W' \right) \, e^{U - W} = \left( e^{U-W} \right)' \;,
\label{eq:Cprime-UW}
\end{equation}
and hence
\begin{equation}
{\hat C} = 1 + \lambda \, e^{-(U- W)} \;,
\label{eq:integr-const-lambda}
\end{equation}
with $\lambda$ a real integration constant.\footnote{Given the relation $C = e^{U - W} + \lambda$ between three of the metric functions, the solution set of the first-order equations \eqref{eqndyonic} is naturally more restricted than the one obtained by looking at the second order equations of motion. In particular, the charged magnetic brane solution of \cite{D'Hoker:2010ij,D'Hoker:2012ih} is not a solution of \eqref{eqndyonic}.}  Inserting this into \eqref{eq:F_rt-sol} gives
\begin{equation}
F_{r t}^A =  - \frac{\lambda}{2} \, e^{- 2 B + V} G^{AB} \hat{q}_B\;.
\label{eq:F_rt-sol3}
\end{equation}
However, the electric field is actually given by
\bea 
(F^A)^{t r} = F_{tr}^A g^{tt} g^{rr} + F_{zr}^A  g^{tz} g^{rr} = \frac12 e^{-2B-W-U-V}G^{AB}\, \hat{q}_B\ , 
\eea
where we used the form of the inverse metric, the third equation of \eqref{gaugefield}, together with the first equation of \eqref{eqndyonic}, and \eqref{eq:F_rt-sol3}. Comparing this with \eqref{flow-e'}, we see that this can also be expressed as
\bea 
(F^A)^{t r} = e^{-2 U - 2V}  ( (e^A)' + A^A_z C')\ .
\eea
Obviously, the electric field is independent of the integration constant $\lambda$ and is non-vanishing whenever some of the charges $\hat{q}_A$ are non-vanishing. 

In contrast, the five-dimensional magnetic field component $(F^A)^{r z}$ does depend on $\lambda$ according to 
\bea 
(F^A)^{rz} = F_{rz}^A g^{rr} g^{zz} + F_{rt}^A  g^{rr} g^{tz} = \frac12 \lambda e^{-2B-W-U-V}G^{AB}\, \hat{q}_B\ .
\eea
Notice, however, that both the electric field and the $rz$-component of the magnetic field are determined by the charges $\hat{q}_A$. As a consequence, the combination $G_{AB} F^A_{MN} F^{B MN}$ vanishes (independently of $\lambda$) for vanishing $P^A$ on any solution of \eqref{eqndyonic}, i.e.\
\bea
G_{AB} F^A_{rt} F^{B rt} = -G_{AB} F^A_{rz} F^{B rz}\ .
\eea

On the other hand, 
inserting \eqref{eq:integr-const-lambda} into the line element \eqref{metricrotation} results in 
\begin{equation}
 ds^2= e^{2 W} \, \lambda \left(\lambda + 2 \, e^{U-W} \right) dt^2 + 2 e^{2W} \, C \, dt\, dz + e^{2W} dz^2
+ e^{2V} \, dr^2+e^{2B}(dx^2+dy^2) \,.
\end{equation}
Thus, we see that the sign and the magnitude of the integration constant $\lambda$ determine the nature of the warped 
line element.  In particular, a vanishing $\lambda$ will give a null-warped metric, i.e.\ $g_{tt} =0$. 

Let us now briefly display the flow equations
for static, 
purely magnetic solutions.  They are obtained by setting $q_A = A_z^A = J = 0$, which results in 
$\hat{q}_A = \hat{J} = C' = 0$, so that the non-vanishing flow equations are 
\begin{eqnarray}\label{magnetic}
(X^A)'&=& \frac23 X^C (gh_Ce^V + G_{CD} P^D e^{V-2B}) X^A - G^{AC} 
(gh_C e^V+  G_{CD} P^D e^{V-2B})\,,\nonumber\\
U'-W'&=&0  \,,\nonumber\\
0&=&B' - \frac12(U' + W') + \frac32 X_A P^A \, e^{V-2B}\,,\nonumber\\
0&=&3 \left(B' + \frac12(U' + W') \right) -  2g X^A h_A e^V+ \frac32 X_A \, P^A \,
 e^{V-2 B} \;.
\end{eqnarray}
These flow equations 
need again to be supplemented by the constraint $h_AP^A=0$. Magnetic supersymmetric $AdS_3 \times \mathbb{R}^2$ 
solutions to these equations were studied in 
\cite{Grover:2008ih,Almuhairi:2010rb,Almuhairi:2011ws,Donos:2011pn}.

Finally, we would like to show that the flow equations \eqref{eqndyonic} follow from a superpotential. To do so, it is convenient to introduce the combinations
\bea \label{phiscalars}
\phi_1 = B-\frac12 (U+W) \quad , \quad \phi_2 = B + \frac12 (U+W) \quad , \quad \phi_3 = U-W\ .
\eea
Using them and introducing the physical scalars $\varphi^i$, the one-dimensional Lagrangian \eqref{lagrange2} takes the form
\bea \label{Lphi}
{\cal L}&=&- e^{2 \phi_2 - V} (\phi_1')^2 + 3 e^{2 \phi_2 - V} (\phi_2')^2- \frac12 e^{2 \phi_2 - V} (\phi_3')^2 \nonumber \\
& & - e^{2 \phi_2 - V} G_{ij}  (\varphi^i)'(\varphi^j)' - e^{\phi_1 + \phi_2 + \phi_3 - V} G_{AB}(A^A_z)'(A^B_z)'  \nonumber \\
&& -e^{-2 \phi_1 + V} G_{AB} P^A P^B - \frac14 e^{-\phi_1 - \phi_2 + \phi_3 + V} G^{AB} \hat{q}_A\hat{q}_B \nonumber \\
&&  -\frac12 e^{-2 \phi_2 + 2 \phi_3 + V} \hat{J}^2 - g^2 e^{2\phi_2 +V}(G^{AB}h_Ah_B-2(h_AX^A)^2) \,,
\eea
where we used \cite{Gunaydin:1984pf, Gunaydin:1984ak}
\bea \label{Gij}
G_{ij} = G_{AB} \partial_i X^A \partial_j X^B\ .
\eea
The advantage of working with the combinations \eqref{phiscalars} is that the sigma-model metric is then block diagonal with 
\bea
&& g_{\phi_1 \phi_1} = e^{2 \phi_2 - V} \quad , \quad g_{\phi_2 \phi_2} = -3 e^{2 \phi_2 - V} \quad , \quad g_{\phi_3 \phi_3} = \frac12 e^{2 \phi_2 - V} \ , \nonumber \\ 
&& g_{ij} =  e^{2 \phi_2 - V} G_{ij} \quad , \quad g_{AB} = e^{\phi_1 + \phi_2 + \phi_3 - V} G_{AB}\ .
\eea
It is now a straightforward exercise to show that the potential $\cal{V}$ of the one-dimensional Lagrangian (i.e.\ the last two lines of \eqref{Lphi}) can be expressed as 
\bea \label{potentialV}
{\cal{V}} = - g^{\phi_1 \phi_1} \left( \frac{\partial \CZ}{\partial \phi_1} \right)^2\!\! - g^{\phi_2 \phi_2} \left( \frac{\partial \CZ}{\partial \phi_2} \right)^2\!\! - g^{\phi_3 \phi_3} \left( \frac{\partial \CZ}{\partial \phi_3} \right)^2\!\! - g^{ij}  \frac{\partial \CZ}{\partial \varphi^i} \frac{\partial \CZ}{\partial \varphi^j} - g^{AB} \frac{\partial \CZ}{\partial A^A_z} \frac{\partial \CZ}{\partial A^B_z}\ ,
\eea
with the superpotential 
\bea \label{CZ}
\CZ = \frac12 e^{\phi_3} \hat J + \frac32 e^{\phi_2 - \phi_1} P^A X_A - e^{2 \phi_2} g h_A X^A\ .
\eea
In doing so, one has to make use of the constraint $h_A P^A = 0$, of \eqref{Gij} and \cite{Gunaydin:1984pf, Gunaydin:1984ak}
\bea \label{Xrelations}
G^{ij} \partial_i X^A \partial_j X^B = G^{AB} - \frac23 X^A X^B \quad , \quad \partial_i X_A = -\frac23 G_{AB} \partial_i X^B  \quad , \quad X_A = \frac23 G_{AB} X^B \ .
\eea
Using the superpotential \eqref{CZ}, it is straightforward to check that the first-order flow equations (in the physical moduli space) can be expressed as\footnote{Note that  \eqref{potentialV} would hold also for any combination of signs in $\CZ = \frac12 e^{\phi_3} \hat J \pm \frac32 e^{\phi_2 - \phi_1} P^A X_A \pm e^{2 \phi_2} g h_A X^A$, but \eqref{flowZ} requires the signs given in \eqref{CZ}.}
\bea \label{flowZ}
\phi_1' = g^{\phi_1 \phi_1} \frac{\partial \CZ}{\partial \phi_1} \ , \quad \phi_2' = g^{\phi_2 \phi_2} \frac{\partial \CZ}{\partial \phi_2} \ , \quad \phi_3' = g^{\phi_3 \phi_3} \frac{\partial \CZ}{\partial \phi_3} \ , \quad (A^A_z)' = g^{AB} \frac{\partial \CZ}{\partial A^B_z}  \ , \quad (\varphi^i)' = g^{ij} \frac{\partial \CZ}{\partial \varphi^j}\ .\nonumber \\
\eea
In order to derive the flow equation for $\varphi^i$, one has to multiply the flow equation for $X^A$ by $G_{AB} \partial_j X^B$ and use \eqref{Gij}, \eqref{Xrelations} and
\bea
X_A \partial_i X^A = 0\ .
\eea
This leads to
\bea
(\varphi^i)' = - G^{ij} e^V \left(g h_A + G_{AC} P^C e^{-2B}\right) \partial_j X^A\ .
\eea
In the absence of fluxes, the superpotential \eqref{CZ} reduces to the one obtained in \cite{Bellucci:2010aq}.


\subsection{Hamiltonian constraint}

Next, we discuss the Hamiltonian constraint and show that it equals the constraint $h_AP^A=0$ that we encountered
in the rewriting of the Lagrangian in terms of first-order flow equations.

The Einstein equations take the form
\begin{eqnarray}
 R_{MN}&=&\frac13 g_{MN}\,g^2\left(G^{AB}h_Ah_B-2(h_AX^A)^2\right)-\frac16 g_{MN}G_{AB}F^A_{KL}F^{B\,KL} \nonumber\\
&& +G_{AB}(X^A)'(X^B)'\delta_{Mr}\delta_{Nr} +G_{AB}F^A_{MK}F^B_{NL}g^{KL}\,.
\end{eqnarray}
There are only five independent equations, namely the ones corresponding to the $tt$-, $rr$-, $xx$-,$zz$- and $tz$-component of the Ricci tensor, which we have displayed in appendix \ref{sec:ricci}.
To obtain the Hamiltonian constraint, we consider the $tt$-component of Einstein's equations.
We use the $rr$-, $xx$-,$zz$- and $tz$-equations to obtain expressions for the second derivatives
$U''$, $C''$, $B''$ and $W''$, which we then insert into the expression for the  $tt$-component.
This yields the following equation, which now only contains first derivatives, 
\begin{eqnarray}
0&=&2B'^2+2U'W'+4B'W'+4B'U'+\frac12e^{2W-2U}C'^2-G_{AB}(X^A)'(X^B)'+G_{AB}F^A_{rt}F^B_{rt}e^{-2U}\nonumber\\
&&+G_{AB}P^AP^Be^{2V-4B}-G_{AB}(A^A_z)'(A^B_z)'(e^{-2W}-e^{-2U}C^2)-2G_{AB}F^A_{rt}(A^B_z)'e^{-2U}C\nonumber\\
&&+g^2 e^{2V}\left(G^{AB}h_Ah_B-2(h_AX^A)^2\right)\,.
\label{hamilt}
\end{eqnarray}
This equation turns out to be equivalent to the $rr$-component of Einstein's equations,
\begin{equation}
R_{rr}-\frac12 g_{rr}R-\frac12g_{rr} {\cal L}_M +\frac{\delta{{ \cal L}_M}}{\delta g^{rr}}=0 \;,
\end{equation}
where ${\cal L}_M$ denotes the matter Lagrangian.

Next, using \eqref{flow-e'}, \eqref{Cprime}, \eqref{eq:Cprime-UW} and the flow equation for $(A^A_z)'$
in \eqref{hamilt}, we obtain
the intermediate result
\begin{eqnarray}
0&=&2B'^2+2U'W'+4B'W'+4B'U'+\frac12(U'-W')^2-G_{AB}(X^A)'(X^B)'\nonumber\\
&&+G_{AB}P^AP^Be^{2V-4B}+g^2 e^{2V}\left(G^{AB}h_Ah_B-2(h_AX^A)^2\right)\,.
\end{eqnarray}
Then, using the first-order flow equation \eqref{eqndyonic} for $(X^A)'$, we get
\begin{eqnarray}\label{constr5d2}
0&=&2B'^2+2U'W'+4B'W'+4B'U'+\frac12(U'-W')^2-\frac43 g^2 e^{2V}(h_AX^A)^2\nonumber\\
&&+\frac23(G_{AB}X^AP^B)^2e^{2V-4B}+\frac43 g e^{2V-2B}(h_AX^A)(G_{AB}X^AP^B)-2ge^{2V-2B}h_AP^A\,.\nonumber\\
&&
\end{eqnarray}
In the next step we use \eqref{eqUBW} as well as the fourth flow equation of 
\eqref{eqndyonic} to obtain
\begin{eqnarray}
0&=&2B'^2+2U'W'+4B'W'+4B'U'+\frac12(U'-W')^2-\frac43 (B'+U'+W')^2\nonumber\\
&&+\frac23\left(-B'+\frac12(U'+W')\right)^2+\frac43(B'+U'+W')\left(-B'+\frac12(U'+W')\right) \nonumber\\
&& -2ge^{2V-2B}h_AP^A\,. \label{hamilt-fin}
\end{eqnarray}
Then, one checks that all the terms containing $B',U'$ and $W'$ cancel out, so that
the on-shell Hamiltonian constraint \eqref{hamilt-fin} reduces to \eqref{hpconstraint}.


\section{Reducing to four dimensions \label{sec:4d5d}}

The five-dimensional stationary solutions to the flow equations \eqref{eqndyonic} may be related to 
a subset of the four-dimensional
static solutions discussed in \cite{Dall'Agata:2010gj,Barisch:2011ui} by performing a reduction on the $z$-direction.
We briefly describe this below.  A detailed check of the matching of the five- and four-dimensional flow
equations is performed in appendix \ref{45dict}.

The five-dimensional solutions are supported by electric fluxes $h_A^{\rm 5d}$, electric charges $q_A^{\rm 5d}$,  
magnetic fields $P^A_{\rm 5d}$, and rotation $J$.\footnote{$J$ generates translations in the $z$-direction. When the coordinate $z$ is compact, $J$ has the interpretation of angular momentum.}  The relevant subset of four-dimensional solutions
is supported by electric fluxes $h_A^{\rm 4d}$, electric charges $Q_I = (Q_0, Q_A)$ and 
magnetic fields $P^A_{\rm 4d}$.

The five-dimensional $N = 2$, $U(1)$ gauged supergravity action \eqref{5d-action-bos} is based on real scalar fields 
$X_{\rm 5d}$ which satisfy the constraint $\frac16 C_{ABC}X_{\rm 5d}^A X^B_{\rm 5d} X^C_{\rm 5D}= 1$ for some constants $C_{ABC}$, while the four-dimensional $N = 2$, $U(1)$  gauged supergravity
action considered in  \cite{Dall'Agata:2010gj,Barisch:2011ui} is based on complex scalar fields $X^I_{\rm 4D}$
with 
a cubic prepotential function
\begin{equation}
F(X_{\rm 4d})=-\frac16 \frac{C_{ABC}X^A_{\rm 4d} X^B_{\rm 4d} X^C_{\rm 4d} }{X^0_{\rm 4d}} \;.
\end{equation}
The four-dimensional physical scalar fields are $z^A = X^A_{\rm 4d}/ X^0_{\rm 4d}$, which we decompose as
$z^A = C^A + i \hat{X}^A$.  

Now we relate the real four-dimensional fields  $(C^A, \hat{X}^A)$ to the fields appearing in the five-dimensional
flow equations.  To do so, we find it convenient to use a different normalization
for the scalar constraint equation, namely
\bea \label{vnorm}
\frac16 C_{ABC}X_{\rm 5d}^A X^B_{\rm 5d} X^C_{\rm 5D}= v\ .
\eea
We will show in the appendix that the matching between the four-dimensional and the five-dimensional flow equations requires to choose $v = \frac12$, a value which was already obtained in \cite{Cardoso:2007rg} when matching the gauge kinetic terms in four and five dimensions. Choosing the normalization \eqref{vnorm} amounts to replacing $C_{ABC}$ by $C_{ABC}/v$, a change that
affects the normalization of the Chern-Simons term in the five-dimensional 
action \eqref{5d-action-bos},  as well as the quantities $\hat{q}_A$ and $\hat{J}$ given in \eqref{hatq} and \eqref{Jhat}, respectively.  On the other hand, if we stick to the definition $X_A^{\rm 5d} = \frac16 C_{ABC} X^B_{\rm 5d}
X^C_{\rm 5d}$, we get  
\bea \label{XA-v}
X_A^{\rm 5d}= \frac{2v}{3} G_{AB} X^B_{\rm 5d}\ ,
\eea 
with $G_{AB}$ given by
\begin{equation}
G_{AB}(X_{\rm 5d})=\frac{1}{v}\left(-\frac12 C_{ABC}X^C_{\rm 5d}+\frac{9}{2v}X_{A}^{\rm 5d}X_{B}^{\rm 5d}\right)\,.
\label{GAB-v}
\end{equation}
Using this normalization, we obtain the following dictionary between the 
four-dimensional
quantities that appeared in \cite{Barisch:2011ui} and the 
five-dimensional 
quantities that enter in  \eqref{eqndyonic},
\begin{eqnarray}
\hat{X}^A &=& e^W \,  X^A_{\rm 5d} \;, \nonumber\\
C^A &=& A_z^A \;, \nonumber\\
h_{A}^{\rm 4d}&=&-h_{A}^{\rm 5d}\,,\nonumber\\
P^A_{\rm 4d}&=&-P^A_{\rm 5d}\,,\nonumber\\
Q_{A}^{\rm 4d}&=&-\frac12 \,q_{A}^{\rm 5d}\,,\nonumber\\
Q_0^{\rm 4d} &=& \frac12 J \;.
\label{4d-5d-rel}
\end{eqnarray}
The five- and four-dimensional line elements are related by
\begin{eqnarray}\label{metricrotation2}
 ds^2_5
 =e^{2 \phi}ds_4^2+e^{- 4 \phi}(dz+Cdt)^2\,,
 \label{rel-line-45}
\end{eqnarray}
where 
\begin{align}
ds^2_4 = - e^{2U_4} dt^2 + e^{-2U_4} dr^2 + e^{-2U_4 + 2 \psi} (dx^2 + dy^2) \;
\end{align}
and
\begin{equation}
2 \phi = - W \;.
\label{phi-W-rel}
\end{equation}
This yields
\begin{eqnarray}
U &=& U_4 + \phi  \;, \nonumber\\
V &=& - U_4 + \phi \;, \nonumber\\
B &=& \psi - U_4 + \phi \;.
\label{rel-warp}
\end{eqnarray}

\section{Solutions \label{sec:ex-sol}}

In the following, we construct solutions to the flow equations \eqref{eqndyonic}.  First we consider exact
solutions with constant scalars $X^A$.  Subsequently we numerically construct solutions with running scalars $X^A$.


\subsection{Solutions with constant scalar fields $X^A$}

We pick $V =0$ in the following.

We will consider two distinct cases.  In the first case, all the magnetic fields $P^A$ are taken to be non-vanishing.
In the second case, we set all the $P^A$ to zero.  Other cases where only some of the $P^A$ are turned on 
are also possible, and their analysis should go along similar lines.


\subsubsection{Taking $P^A \neq 0, \, \hat{q}_A=0, \hat{J} \neq 0$}
\label{sec:ex-sol_nonvanP}

Here we consider the case when all the $P^A$ are turned on. Demanding $X^A = {\rm constant}$ yields
\begin{equation}
X^C (gh_C  + G_{CD} P^D e^{-2B}) X_A = 
gh_A +  G_{AB} P^B e^{-2B}\;.
\label{X-const-value}
\end{equation}
Observe that $G_{AB}$ is constant, and so is $B$. We set $B=0$ in the following, which can always be achieved by rescaling $x$ and $y$.  Combining \eqref{X-const-value}
with \eqref{B-cond}, we express the magnetic fields $P^A$ in terms of $h_A$ and $X^A$ as 
\begin{equation}
P^A = - g \, G^{AB} \left(h_B - \frac32 \left(h_C \, X^C \right) X_B \right) \;.
\label{P-g-X}
\end{equation}
This relation generically fixes the scalars $X^A = X^A (h_B, P^B)$ in terms of the fluxes and magnetic fields, as we will see in the explicit examples of sec.\ \ref{sec:non-constscalars}. Contracting \eqref{P-g-X} with $h_A$ and using the constraint
\eqref{hpconstraint} we obtain 
\begin{equation}
G^{AB} \, h_A h_B = \left( h_A \, X^A \right)^2 
\label{hh-hX}
\end{equation}
as well as 
\begin{equation}
G_{AB} P^A P^B = \frac12 g^2 \left( h_A \, X^A \right)^2 \;.
\label{PP-hX}
\end{equation}
Observe that \eqref{hh-hX} together with $h_A \, X^A =0$ would imply $h_A =0$.  Thus, in the following, we take $h_A \, X^A \neq 0$.

We obtain from \eqref{eqUBW},
\begin{equation}
U + W = g h_A X^A \, (r - r_0) \;,
\label{upw-sol}
\end{equation}
where $r_0$ denotes an integration constant.  
Inserting this into the third equation of \eqref{eqndyonic} gives
\begin{equation}
\left( e^{-(U - W)} \right)' = - \hat{J} \, e^{ g h_A X^A (r_0 -r)}  \;.
\label{U-Wtau}
\end{equation}

Next we set $\hat{q}_A =0$, so that the $A_z^A$ take constant values.  These are determined by
\begin{equation}
q_A + C_{ABC}  \, P^B \, A_z^C = 0 \;.
\label{value-Az}
\end{equation}
Defining $C_{AB} = C_{ABC} P^C$, this is solved by
\begin{equation}
A_z^A = - C^{AB} \, q_B \;,
\end{equation}
where $C^{AB} C_{BC} = \delta^A{}_C$.  Here, we assumed that $C_{AB}$ is invertible, which generically is the 
case when all the $P^A$ are turned on.  

For constant $A_z^A$, $\hat{J}$ is also constant, and we can solve \eqref{U-Wtau}.  
Taking $h_A X^A \neq 0$, we get
\begin{equation} \label{UWb}
e^{-(U - W)} = \frac{\hat{J} \, e^{g h_A X^A (r_0-r)} }{g h_A X^A}  + b\;,
\end{equation}
where $b$ denotes an integration constant.
Combining this result with \eqref{upw-sol} gives
\begin{equation}
e^{2W} = \frac{\hat{J} }{g h_A X^A}  + b  \, 
e^{g h_A X^A \, (r-r_0)} 
\end{equation}
as well as 
\begin{equation}
e^{- 2 U } = \frac{\hat{J} \, e^{2g h_A X^A (r_0 - r)} }{g h_A X^A} 
 + b \, e^{g h_A X^A \, (r_0 -r )}  \;.
\end{equation}
To bring these expressions into a more palatable form, we introduce
a new radial variable
\begin{equation} \label{taur}
\tau = \alpha \, e^{g h_A X^A \, (r-r_0)} 
\end{equation}
with $\tau \geq 0$ and 
\begin{equation} \label{definealpha}
\alpha =  g h_A X^A > 0 \;.
\end{equation}
(If $g h_A X^A <0$ we have $\tau \leq 0$.)
Then (assuming $\hat J \geq 0$)
\begin{eqnarray} \label{WUC}
e^{2W} &=& \alpha^{-1} \left[ \hat{J}  + b  \, \, \tau \right] \;,\nonumber\\
e^{- 2 U } &=& \alpha \left[ \frac{\hat{J}}{\tau^2}
 + \frac{b}{\tau} \right] \;,\nonumber\\
 C &=& \frac{\tau}{\hat{J} + b \, \tau} + \lambda \;,
\end{eqnarray}
and the associated line element reads
\begin{eqnarray}
ds^2 &=& - \alpha^{-1}
\frac{\tau^2}{\hat{J}\,+\,b\,\tau}\, dt^2 
 +  \frac{\alpha^{-2} }{ \tau^2} \, d\tau^2 \nonumber\\
 &&
 + \alpha^{-1} \, \left(\,\hat{J}\,+\,b\,\tau\,\right)
  \left( \,dz \,+\, \left[\frac{\tau}{\hat{J}\,+\,b\,\tau}+\lambda\right]\, dt \right)^2 \nonumber\\
 && + \left(dx^2 + dy^2 \right) \;,
 \end{eqnarray}
Now we notice that for 
\begin{equation} \label{cE}
\lambda=-\frac{1}{b}\ \  {\rm and} \ \ b = 4 \alpha^{-3}  > 0 
\end{equation} 
and assuming $z$ to be compact, this is nothing but the metric of the extremal BTZ black hole in $AdS_3$ times $\mathbb{R}^2$, so that the space time is asymptotically $AdS_3\,\times\,\mathbb{R}^2$. This can be made manifest by the coordinate redefinitions
\begin{eqnarray}
\tau = \rho^2 - \frac{\hat{J}}{b}\ , \quad z = \frac{l}{b} \, \phi \ ,
 \end{eqnarray}
 where $l^2 = \alpha \, b = 4/( g h_A X^A)^2$.  Introducing
 \begin{eqnarray}
 j = \frac{2 \hat{J}}{b l} \;,
 \end{eqnarray}
 the line element becomes
 \begin{eqnarray}
ds^2 = - \left(\frac{\rho}{l}- \frac{j}{2\rho} \right)^2\, dt^2 + \left(\frac{\rho}{l}- \frac{j}{2\rho} \right)^{-2}
\, d \rho^2 + 
 \rho^2  \left( \,d \phi \,-\, \frac{j}{2 \rho^2} \, dt \right)^2 + \left(dx^2 + dy^2 \right) .
 \label{btzr2}
 \end{eqnarray}
This describes an extremal BTZ black hole with  angular momentum $j$ and mass $M = j/l$ \cite{Banados:1992wn}, where
$l$ denotes the radius of $AdS_3$.  The horizon is at $\rho_+^2 = j l /2 = \hat{J}/b$, which corresponds to $\tau =0$.
The entropy of the BTZ black hole (and hence the entropy density of the extremal BTZ $\times \mathbb{R}^2$ solution \eqref{btzr2}) is 
\begin{equation}
{\cal S}_{\rm BTZ}  = \frac{2 \pi \, \rho_+}{4} = \pi \frac{\sqrt{\hat{J}}}{4} \, \alpha^{3/2} \;.
\label{entrobtz}
\end{equation}
Observe that $\alpha$ is determined in terms of the fluxes $h_A$ and the $P^A$ through \eqref{P-g-X}, 
and so it is independent of $J$ and $q_A$.

In deriving the above solution, we have assumed that all the $P^A$ are turned on so as to ensure
the invertibility of the matrix $C_{AB}$. In this generic case, the constant values of the scalar fields 
$X^A$ and $A_z^A$ are entirely determined in terms of the $h_A, P^A$ and $q_A$.  When switching off some of the $P^A$,
some of the $A_z^A$ may be allowed to have arbitrary constant values, but these are expected not to contribute
to the entropy density.

The BTZ$\times \mathbb{R}^2$-solution given above can be found in any $N = 2$, $U(1)$ gauged supergravity model. 
This can also be inferred as follows.  Setting $B' = (X^A)' = (A_z^A)' = \hat{q}_A = 0$ in the Lagrangian \eqref{lagrange2}
and using the relations \eqref{hh-hX} and \eqref{PP-hX} yields a one-dimensional Lagrangian that descends from 
a three-dimensional Lagrangian describing
Einstein gravity in the presence of an anti-de Sitter cosmological constant $\Lambda = - 1/l^2$ determined by the flux potential
($4/l^2 = (g h_A X^A)^2$).
As is well-known, the associated three-dimensional equations of motion allow for 
extremal BTZ black hole solutions with rotation.  As shown in \cite{Grover:2008ih}, the near-horizon geometry
of the BTZ$\times \mathbb{R}^2$-solution (which is supported by the magnetic fields \eqref{P-g-X}) preserves
half of the supersymmetry.

The entropy of the BTZ black hole solution depends on $\hat{J}$, which
takes the form
\begin{equation}
J + \frac12 \, C^{AB} q_A q_B \;.
\label{spectral-inv_BTZ}
\end{equation}
This combination is invariant under the transformation
\begin{equation}
J \rightarrow \hat{J} \;\;\;,\;\;\; q_A \rightarrow \hat{q}_A \;,
\label{spectral}
\end{equation}
with $\hat{J}$ and $\hat{q}_A$ given in \eqref{Jhat} and \eqref{hatq}, respectively.
In the absence of fluxes, this transformation
 is called spectral flow transformation and can be understood as follows from the supergravity perspective
 \cite{deBoer:2006vg}.
The rewriting of the five-dimensional Lagrangian in terms of first-order flow equations makes use of the
combinations $\hat{q}_A$ and $\hat{J}$.  These combinations have their origin in the presence of the gauge
Chern-Simons term.
When the $A_z^A$ are constant, the shifts
$q_A \rightarrow \hat{q}_A$ and $J \rightarrow \hat{J}$ take the form of shifts induced by a large gauge
transformation of $A_z^A$, i.e. $A^A \rightarrow A^A + k^A$, where $k^A$ denotes a closed one-form.  
These transformations
constitute a symmetry of string theory, and this 
implies that the entropy of a black hole should be invariant under spectral flow.  It must
therefore depend on the combination \eqref{spectral-inv_BTZ}.
In the presence of fluxes, we find that the BTZ $\times \mathbb{R}^2$-solution \eqref{btzr2} respects 
the spectral flow transformation \eqref{spectral}. 

The three-dimensional extremal BTZ black hole geometry, resulting from dimensionally reducing the black brane solution
\eqref{btzr2}
on $\mathbb{R}^2$, is a state in the two-dimensional CFT dual to the asymptotic $AdS_3$, with left-moving central charge $c\,=\,\frac{3 l}{2 G_3}$ and $L_0\,-\frac{c}{24}\,=\,\frac{\rho^2_+}{4\,G_3\,l}$, as well as $\tilde{L}_0 - \frac{\tilde c}{24} = 0$, cf.\ \cite{Kraus:2006wn}. Hence, the large charge leading term in the entropy of the black hole is given by the Ramanujan-Hardy-Cardy formula for the dual CFT,
\begin{equation}
{\cal S}_{\rm BTZ}\,=\, 2\,\pi\,\sqrt{\frac{c}{6}\, \left(L_0\,-\,\frac{c}{24}\right)}\ .
\end{equation}
This is exactly equal to the Bekenstein-Hawking entropy \eqref{entrobtz} computed above (in units of $G_3=1$), and
can be regarded as a microscopic computation of the bulk black brane entropy from the holographic dual CFT.


\subsubsection{Taking $P^A =0, q_A =0, \hat{J} \neq 0$}

Now we consider the case when all the $P^A$ vanish.  Then, \eqref{X-const-value} reduces to 
\begin{equation}
\left( h_C X^C \right) X_A = 
h_A \;,
\label{X-const-value-P0}
\end{equation}
which determines the constants $X^A$ in terms of the fluxes $h_A$.
Contracting \eqref{X-const-value-P0}
with $G^{AB} h_B$ yields
\begin{equation}
G^{AB}\, h_A h_B = \frac23 \left(h_A \, X^A \right)^2 \;.
\label{hh-hX-P0}
\end{equation}
Thus we take $h_A \, X^A \neq 0$ in the following, since otherwise $h_A =0$.

Combining the fourth equation of \eqref{eqndyonic} with \eqref{eqUBW} results in
\begin{equation}
B' = \frac13 g h_A X^A \;,
\end{equation}
which can be readily integrated to give
\begin{equation}
e^B = e^{\beta} \, e^{\frac13 g h_A X^A \, r} \;,
\end{equation}
where $\beta$ denotes an integration constant which we set to zero.  The combination $U+W$ is given by
\begin{equation}
U + W = 2 B + u \;,
\label{upwb}
\end{equation}
where $u$ denotes an integration constant which we also set to zero.

The flow equation for $U-W$ reads
\begin{equation}
\left( e^{-(U-W)} \right)'= - \hat{J} \, e^{-4B} = - \hat{J} \, 
e^{- \frac43 g h_A X^A \, r} \;.
\label{diffU-W}
\end{equation}

Next, let us consider the flow equation for $A_z^A$, 
\begin{equation}
\left( A_z^A \right)' = - \tfrac12 G^{AC} \, q_C \, e^{-2B} \;.
\end{equation}
A non-vanishing $q_A$ yields a running scalar field
$A_z^A \sim G^{AB} q_B \, e^{- \frac23 g h_A X^A \, r} $.
In the chosen coordinates, the line element can only have a throat at $|r |= \infty$.
At either of these points, either the area element $e^B$ or $A_z^A$ blows up. If we demand that both $e^B$ and $A_z^A$ stay finite at the horizon, we are thus led to take $A_z^A$ to be constant, which can be obtained by setting $q_A = 0$.
Therefore, we set
$q_A =0$ in the following.
This implies
that $\hat{J}$ is constant, which we take to be non-vanishing.

Taking 
$h_A X^A \neq 0$, \eqref{diffU-W} is solved by
\begin{equation}
e^{-(U-W)} = \frac34 \frac{\hat{J} }{g h_A X^A} \, e^{- \frac43 g h_A X^A \, r} + \gamma \;,
\end{equation}
where $\gamma$ denotes an integration constant.
Using \eqref{upwb}, this results in 
\begin{eqnarray}
e^{2W} &=& \frac34 \frac{\hat{J} }{g h_A X^A} \, e^{- \tfrac23 g h_A X^A \, r} + \gamma 
\,
e^{\tfrac23 g h_A X^A \, r}\;,
\nonumber\\
e^{-2 U} &=& \frac34 \frac{\hat{J} }{g h_A X^A} \, e^{- 2 g h_A X^A \, r} + \gamma \, 
\, e^{- \tfrac23 g h_A X^A \, r}\;.
\end{eqnarray}
Redefining the radial coordinate,
\begin{equation}
\tau = e^{\tfrac13 g h_A X^A \, r}  \;\;\;,\;\;\; \tau \geq 0 \;, 
\end{equation}
yields
\begin{eqnarray}
e^B &=& \tau \;, \nonumber\\
e^{2W} &=& \frac34 \frac{\hat{J}} {g h_A X^A} \, \tau^{-2} + \gamma \, \tau^2 \;,
\nonumber\\
e^{-2U} &=&  \frac34 \frac{\hat{J} }{g h_A X^A} \, \tau^{-6} + \gamma \, \tau^{-2} \;,
\nonumber\\
C &=& \frac{\tau^4}{\frac34 \frac{\hat{J} }{g h_A X^A}  + \gamma \, \tau^4 } + \lambda \;.
\end{eqnarray}
In the chosen coordinates, the line element reads
\begin{equation}
 ds^2=-e^{2U(\tau)} dt^2+ (\tfrac13 g h_A X^A)^{-2} \left(\frac{d\tau}{\tau}\right)^2
  +e^{2B(\tau)}(dx^2+dy^2)+e^{2W(\tau)}(dz+C(\tau)dt)^2\,.
  \label{nernst-sol}
\end{equation}
It exhibits a throat as $\tau \rightarrow 0$.  In the following we set $\lambda =0$, and we take
$\hat{J}/(g h_A X^A) >0$.

In the throat region, the terms proportional to $\gamma$ do not contribute (recall that we are taking $\hat{J}$ to be non-vanishing)
and 
the line element becomes
\begin{equation}
 ds^2=- \tau^6 dt^2+ (\tfrac13 g h_A X^A)^{-2} \left(\frac{d\tau}{\tau}\right)^2
  + \tau^2 (dx^2+dy^2)+ \tau^{-2} (dz+ \tau^4  dt)^2\,,
  \label{interm-lin}
\end{equation}
where we rescaled the coordinates by various constant factors. 
Then, performing the coordinate transformation
\begin{equation}
\tilde{\tau} = \tau^3 \;\;\;,\;\; \tilde{t} = 3 t \;,
\end{equation}
and setting $(\tfrac13 g h_A X^A)^{2} =1 $ for convenience, 
the line element becomes
\begin{eqnarray}
 ds^2 &=& \frac19 \left( - \tilde{\tau}^2 d\tilde{t}^2+ \left(\frac{d\tilde{\tau}}{\tilde{\tau}}\right)^2 \right)
  + \tilde{\tau}^{2/3} (dx^2+dy^2)+ \tilde{\tau}^{-2/3} \left(dz+ \tfrac13 \tilde{\tau}^{4/3} d\tilde{t}\right)^2
  \nonumber\\
  &=&  \frac23 \, \tilde{\tau}^{2/3} d\tilde{t} dz + \tilde{\tau}^{-2/3} dz^2 + 
    \frac19 \left(\frac{d\tilde{\tau}}{\tilde{\tau}}\right)^2 
  + \tilde{\tau}^{2/3} (dx^2+dy^2) \;,
  \label{line-nernst-warp}
  \end{eqnarray}
which describes a null-warped throat.  The entropy density vanishes, ${\cal S} \sim e^{2B + W}|_{\tau=0} = 0$.

In the limit $\tau \rightarrow \infty$, on the other hand, there are two distinct cases.  When $\gamma \neq 0$
(we take $\gamma >0$),
\begin{eqnarray}
e^B &=& \tau \;, \nonumber\\
e^{2W} &\approx& \gamma \, \tau^2 \;,
\nonumber\\
e^{-2U} &\approx&   \gamma \, \tau^{-2} \;,
\nonumber\\
C &\approx& \gamma^{-1}  \;,
\end{eqnarray}
and the line element becomes
\begin{equation}
 ds^2=- \tau^2 dt^2+ (\tfrac13 g h_A X^A)^{-2} \left(\frac{d\tau}{\tau}\right)^2
  + \tau^2 (dx^2+dy^2)+ \tau^{2} (dz + dt)^2\,,
\end{equation}
where we rescaled the coordinates.  Observe that this
describes a patch of $AdS_5$.  

The other case corresponds to setting $\gamma =0$, in which case the behavior at $\tau \rightarrow \infty$
is determined by
\begin{eqnarray}
e^B &=&  \tau \;, \nonumber\\
e^{2W} &=& \frac34\, \frac{\hat{J} }{g h_A X^A} \, \tau^{-2} \;,
\nonumber\\
e^{-2U} &=&  \frac34\, \frac{\hat{J} }{g h_A X^A} \, \tau^{-6} \;,
\nonumber\\
C &=& \frac{4}{3} \,\frac{ g h_A X^A}{\hat{J}}
 \, \tau^4 \;,
\end{eqnarray} 
and the associated line element is again of the form \eqref{interm-lin} and \eqref{line-nernst-warp}.

Thus, we conclude that the solution \eqref{nernst-sol} with $\gamma \neq 0$
describes a solution that interpolates between
$AdS_5$ and a null-warped Nernst throat at the horizon with vanishing entropy, in which
all the scalar fields are kept constant. This is a purely gravitational stationary solution that is supported by electric fluxes $h_A$.
It is an example of a Nernst brane (i.e.\ a solution with 
vanishing entropy density), and  
is the five-dimensional counterpart of the four-dimensional Nernst solution constructed in \cite{Barisch:2011ui}.
Nernst solutions suffer from the problem of divergent tidal forces.  These may get cured by quantum or stringy effects
\cite{Harrison:2012vy}.


\subsection{Solutions with non-constant scalar fields $X^A$}
\label{sec:non-constscalars}

Here we present numerical solutions that are supported by non-constant scalar fields $X^A$ and that interpolate between a near horizon solution of the type discussed above in sec.\ \ref{sec:ex-sol_nonvanP} and an asymptotic $AdS_5$-region with metric
\bea \label{adsasympt}
 ds^2=-e^{2r} dt^2+dr^2+e^{2r}(dx^2+dy^2+dz^2)\ ,
\eea
i.e.\ the metric functions $U(r), B(r)$ and $W(r)$ in \eqref{metricrotation} all asymptote to the linear function $r$ and $C$ becomes $0$ (or constant, since a constant $C$ can be removed by a redefinition of the $z$-variable). To be concrete, we work within the STU-model. Within this model, a solution with a single running scalar and with $\hat J=0$ was already given in sec.\ 2.3 of \cite{Donos:2011pn}. In order to facilitate the comparison with their results, we will work with physical scalars in this section, i.e.\ we solve the constraint $X^1 X^2 X^3 = 1$ via
\bea
X^1 = e^{-\frac{1}{\sqrt{6}} \phi^1 - \frac{1}{\sqrt{2}} \phi^2}\ , \quad X^2 = e^{-\frac{1}{\sqrt{6}} \phi^1 + \frac{1}{\sqrt{2}} \phi^2}\ , \quad X^3 = e^{\frac{2}{\sqrt{6}} \phi^1}\ .
\eea
In an asymptotically $AdS_5$-spacetime the two scalars $\phi^i$ have a leading order expansion
\bea \label{sourcevev}
\phi^i = a_i r e^{- 2 r} + b_i e^{- 2 r} + {\cal O} (e^{-4 r})\ , \quad i = 1,2\ ,
\eea
i.e.\ they both correspond to dimension $2$ operators of the dual field theory and $a_i$ and $b_i$ correspond to the sources and 1-point functions, respectively. 

For all the following numerical solutions, we choose 
\bea
g = 1\ , h_1=h_2=h_3=1\ .
\eea


\subsubsection{Solution with a single running scalar}

Let us first consider the case with a single running scalar field and with vanishing $\hat J$. For concreteness, we choose the magnetic fields 
\bea
P^1 = P^2 = 4^{1/3}\ , \quad P^3 = - 2 \cdot 4^{1/3}\ ,
\eea
so as to satisfy the constraint \eqref{hpconstraint}. One could choose a different overall normalization for the magnetic fields $P^A$ by rescaling the $x$ and $y$ coordinates. This, on the other hand, would also imply a rescaling of $e^B$ and, thus, we would not have $B=0$ anymore, as was assumed in sec.\  \ref{sec:ex-sol_nonvanP}. Solving \eqref{P-g-X} for these values of $P^A$ leads to 
\bea
X^1_{(0)} = X^2_{(0)} = 4^{-1/3}\ , \quad X^3_{(0)} = 4^{2/3}\ ,
\eea
where we added the subscript $(0)$ in order to distinguish the constant near horizon values from the full, non-constant solution to be discussed momentarily. These values lead to 
\bea
\phi^1_{(0)} \approx 1.132\ , \quad \phi^2_{(0)} = 0\ , \quad \alpha = 3\cdot 2^{1/3}\approx 3.780 \ ,
\eea
with $\alpha$ defined in \eqref{definealpha}. Finally, using \eqref{taur}, \eqref{WUC} and \eqref{cE}, we obtain
\bea \label{U0W0C0}
U_{(0)} = -\frac12 \ln \left(4 \alpha^{-3}\right) + \frac{\alpha}{2} r \quad , \quad W_{(0)} = \frac12 \ln \left(4 \alpha^{-3}\right) + \frac{\alpha}{2} r \quad , \quad  C_{(0)} = 0. 
\eea

In order to obtain an interpolating solution with an asymptotic $AdS_5$-region, we slightly perturb around this solution. In particuar, we make the following ansatz for small $r$
\bea \label{perturbansatz}
&& X^1 = X^1_{(0)} + c_1 e^{c_2 r}\ , \quad X^2 = X^2_{(0)} + c_3 e^{c_4 r}\ , \nonumber \\ 
&& B = \delta B \ , \quad U = U_{(0)} + \delta U\ , \quad W = W_{(0)} + \delta W
\eea
with $c_2, c_4 > 0$. Plugging this into the flow equations \eqref{eqndyonic}, we obtain the following conditions:
\bea \label{perturb}
&& c_1 = c_3 \ , \quad c_2 = c_4 = 2^{-2/3} \left( \sqrt{33} - 1 \right) \approx 2.989\ , \nonumber \\
&& \delta U = \delta W = -\frac12 \frac{c_3 (2^{2/3} c_2 + 14)}{2} e^{c_2 r} \ , \quad \delta B =  2^{-2/3} c_3 (3 \cdot 2^{1/3} + c_2)  e^{c_2 r}\ .
\eea
\begin{figure}[t]
\centering \includegraphics[width=0.5\textwidth]{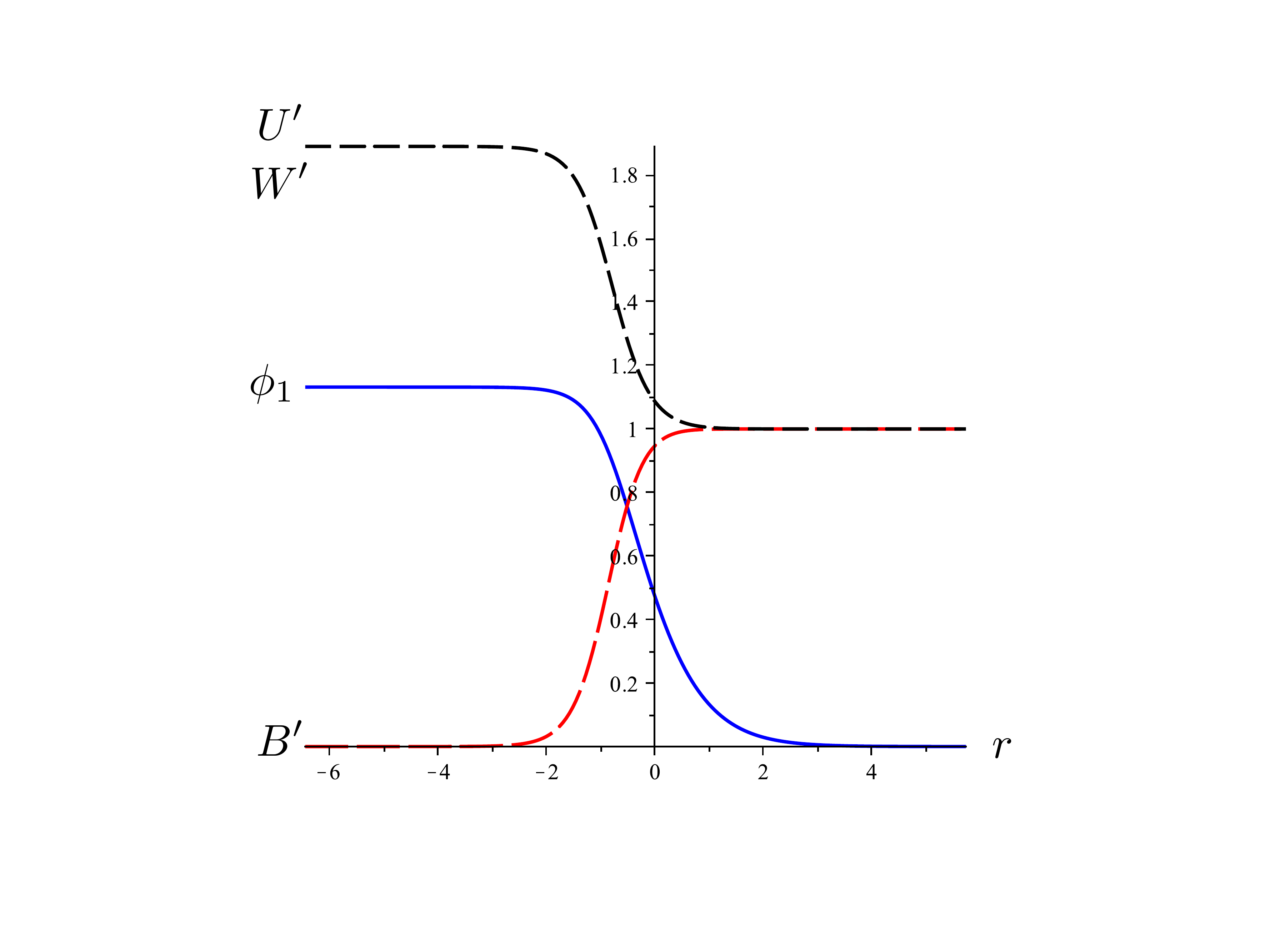}  \includegraphics[width=0.49\textwidth]{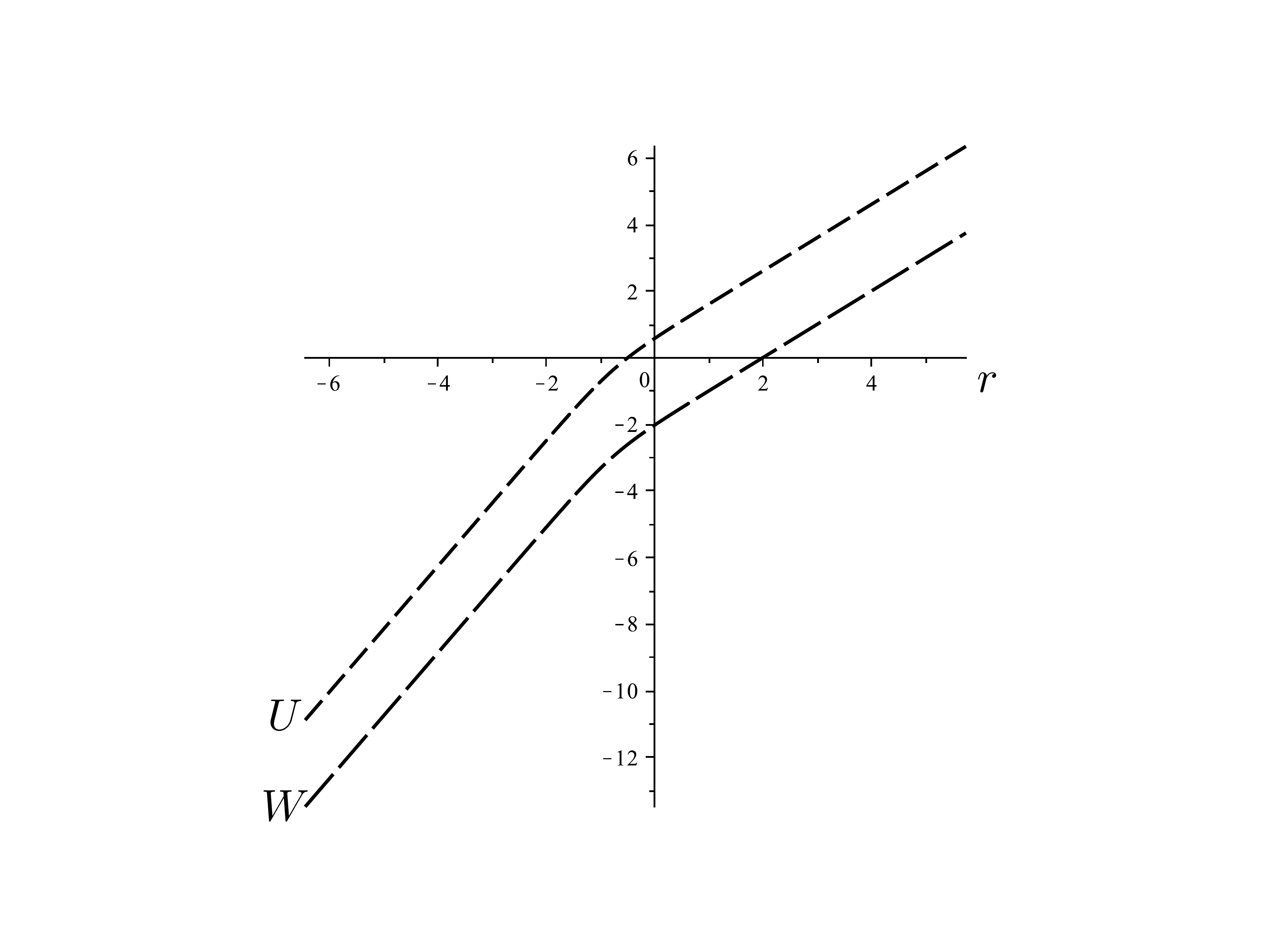} \caption{The case of one scalar with vanishing $\hat J$.\label{1scalar}}
\end{figure}
The constant $c_3$ is undetermined and sets the value of the source and 1-point function of $\phi^1$, cf.\ \eqref{sourcevev}. We will see this explicitly in the example in sec.\ \ref{sec:Jneq0} below. Using \eqref{perturb}, we can find the initial conditions needed to solve \eqref{eqndyonic} numerically. In practice it is most convenient to solve \eqref{eqndyonic} in the $\tau$ variable \eqref{taur}, as the horizon is at $\tau = 0$. This allows to set the initial conditions for instance at $\tau = 10^{-13}$ and then integrate outwards. Doing so and choosing $c_3 = 1$, we obtain the result depicted in figure \ref{1scalar} (note that the plot makes use of the $r$-variable, i.e.\ the primes denote derivatives with respect to $r$ as before). Moreover, $C \equiv 0$. Even though the functions $U$ and $W$ always have the same derivative, they differ by a shift, as can be seen in the right part of figure \ref{1scalar}. This is due to the fact that we chose $b= 4 \alpha^{-3}$ according to \eqref{cE}, instead of $b=1$, with $b$ being introduced in \eqref{UWb}. This solution is very similar to the one discussed in sec.\ 2.3 of \cite{Donos:2011pn} and it is obvious that the metric asymptotically becomes of the form \eqref{adsasympt}.


\subsubsection{Solution with two running scalars, $\hat J = 0$}

\begin{figure}[t]
\centering \includegraphics[width=0.5\textwidth]{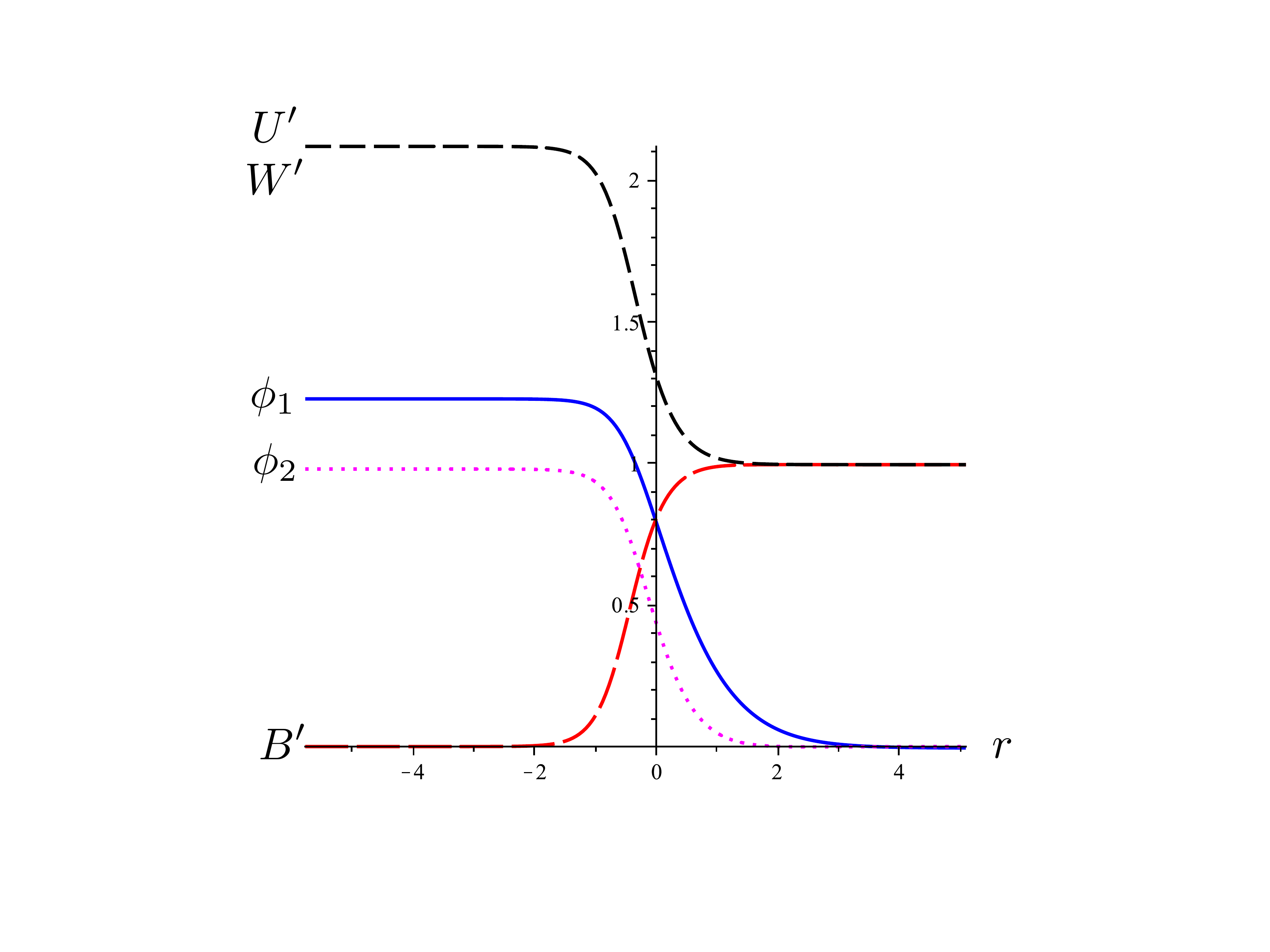}  \caption{The case of two scalars with vanishing $\hat J$.\label{2scalarsnoJ}}
\end{figure}

We now turn to the case of two non-trivial scalars, first still with vanishing $\hat J$ and then with $\hat J \neq 0$ in the next subsection. In all cases we choose 
\bea \label{P1P2P3}
P^1 = \left( \frac43 \right)^{1/3}\ , \quad P^2 = 2 \cdot \left( \frac43 \right)^{1/3}\ , \quad P^3 = - 3 \cdot \left( \frac43 \right)^{1/3}\ .
\eea
Again, the overall normalization of the $P^A$ is imposed on us by demanding $B=0$ in the near-horizon region. This time solving \eqref{P-g-X} leads to 
\bea \label{XA0s}
X^1_{(0)} = \frac14 \left( \frac43 \right)^{2/3}\ , \quad X^2_{(0)} =  \left( \frac43 \right)^{2/3}\ , \quad X^3_{(0)} = \frac32 6^{1/3}\ ,
\eea
which implies
\bea \label{phi1phi2alpha}
\phi^1_{(0)} \approx 1.228\ , \quad \phi^2_{(0)} \approx 0.980\ , \quad \alpha = \frac76 \cdot 4^{2/3} \cdot 3^{1/3}\approx 4.240 \ .
\eea
The functions $U_{(0)}$, $W_{(0)}$ and $C_{(0)}$ are again given by \eqref{U0W0C0}, now using the value of $\alpha$ given in \eqref{phi1phi2alpha}.

Perturbing around the near-horizon solution utilizes the same ansatz as in \eqref{perturbansatz}. This time, we obtain the conditions
\bea \label{perturb2}
&& c_1 = -\frac{c_3}{8} \frac{41 \cdot 6^{1/3} + 33 c_2}{29\cdot 6^{1/3}-3 c_2} \ , \quad c_2 = c_4 \approx 3.694\ , \nonumber \\
&& \delta U = \delta W = \frac{1}{16} \frac{c_3 (21\cdot 6^{2/3} c_2^2 + 588 c_2 - 178\cdot 6^{1/3})}{c_2 (-3 c_2 + 29 \cdot 6^{1/3})} e^{c_2 r} \ ,  \nonumber \\
&& \delta B =  \frac{ 6^{1/3}}{16} \frac{c_3 (-27 c_2^2 - 42 \cdot 6^{1/3} c_2 + 49 \cdot 6^{2/3})}{-3 c_2 + 29 \cdot 6^{1/3}} e^{c_2 r}\ .
\eea
Again, the parameter $c_3$ determines the sources and 1-point functions of the scalars $\phi^1$ and $\phi^2$. 

Using \eqref{perturb2} in \eqref{perturbansatz} we obtain the initial conditions to solve the flow equations numerically. We do not find any solution for $c_3=1$, but inverting the sign, i.e.\ choosing $c_3=-1$, leads to the solution depicted in figure \ref{2scalarsnoJ}, which in addition has $C \equiv 0$ and which is asymptotically $AdS_5$. 


\subsubsection{Solutions with two running scalars, $\hat J \neq 0$}
\label{sec:Jneq0}

\begin{figure}[t]
\centering \includegraphics[width=0.45\textwidth]{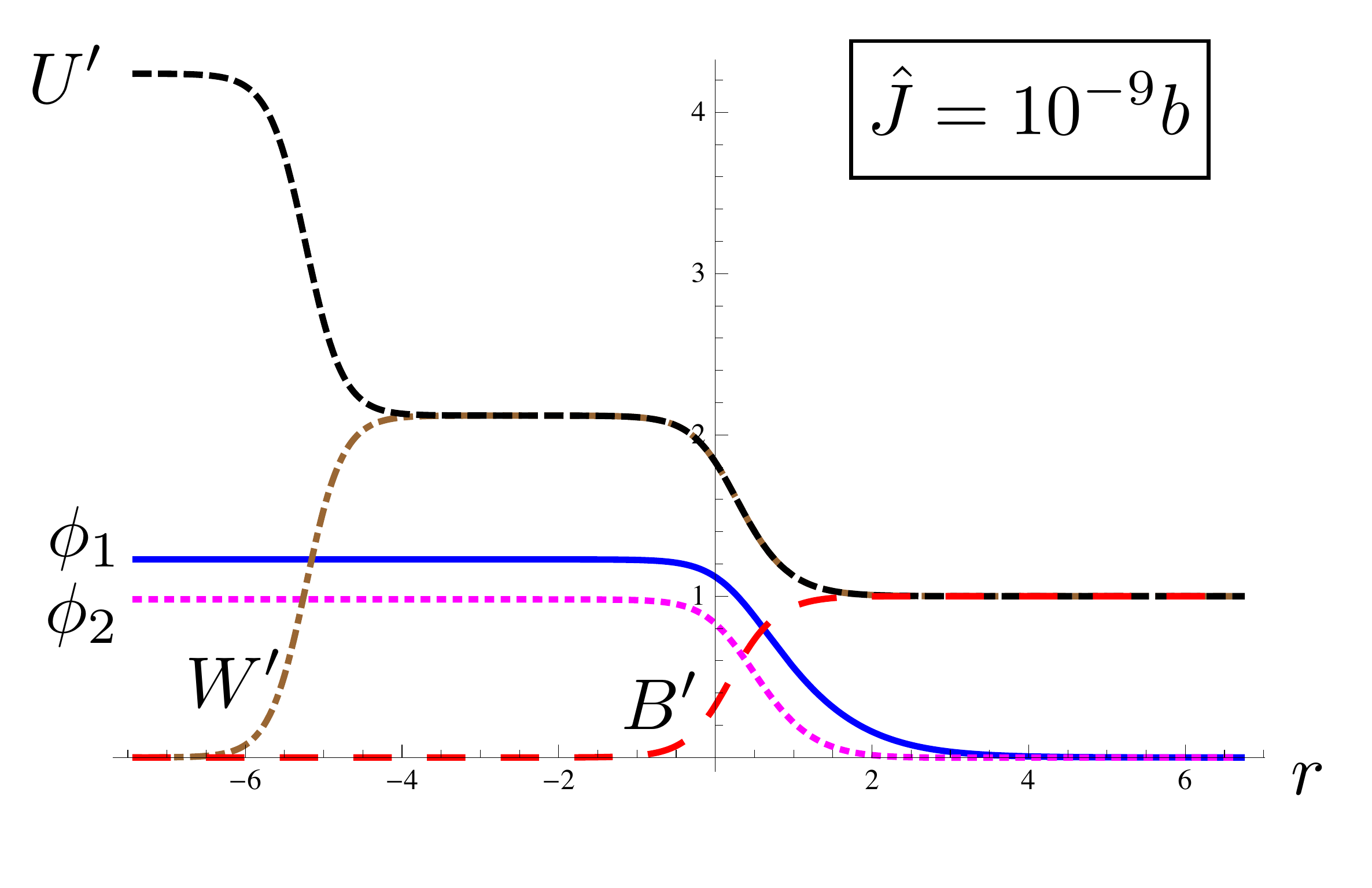} \qquad \includegraphics[width=0.45\textwidth]{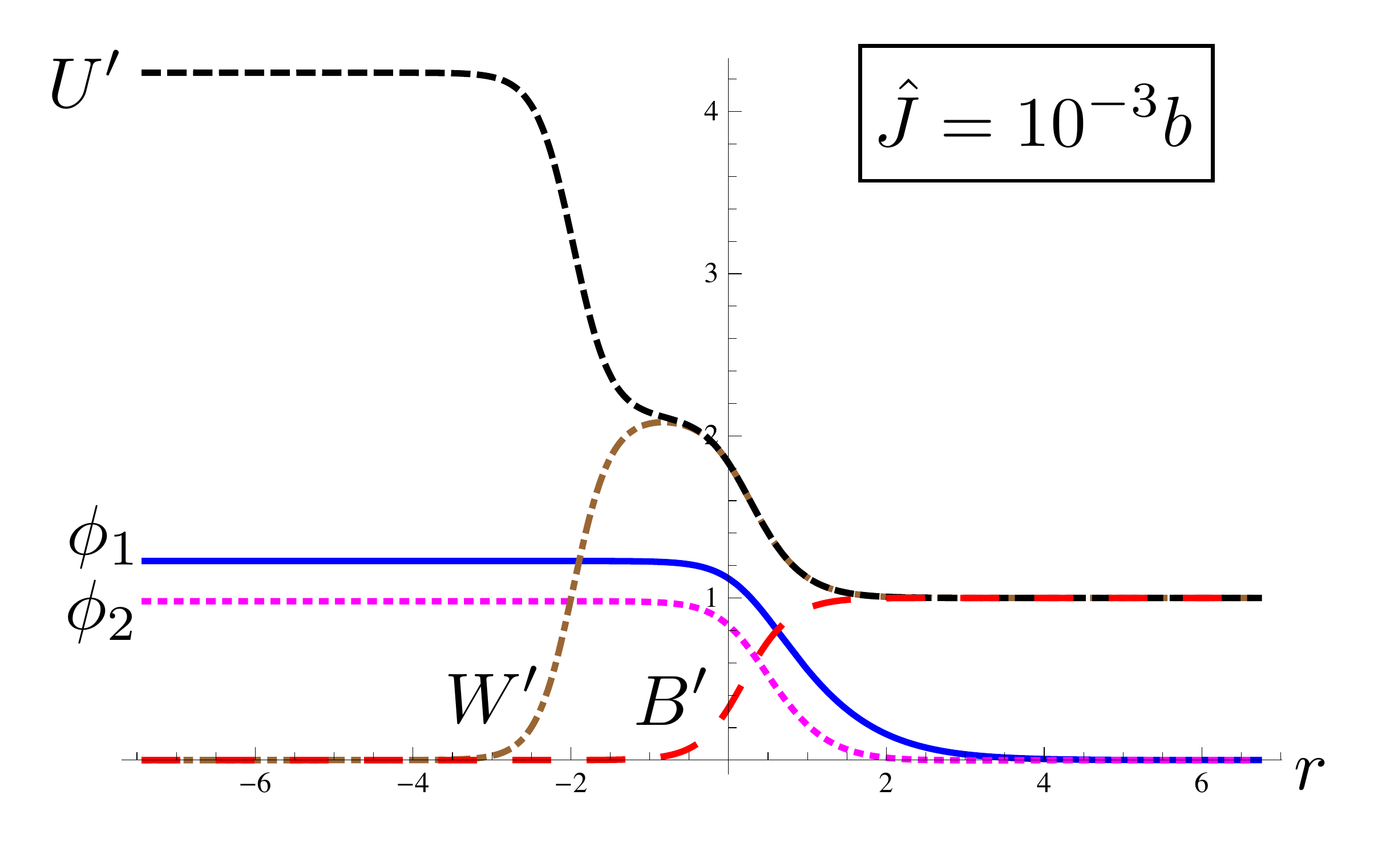}  \\\vspace{1cm} \includegraphics[width=0.45\textwidth]{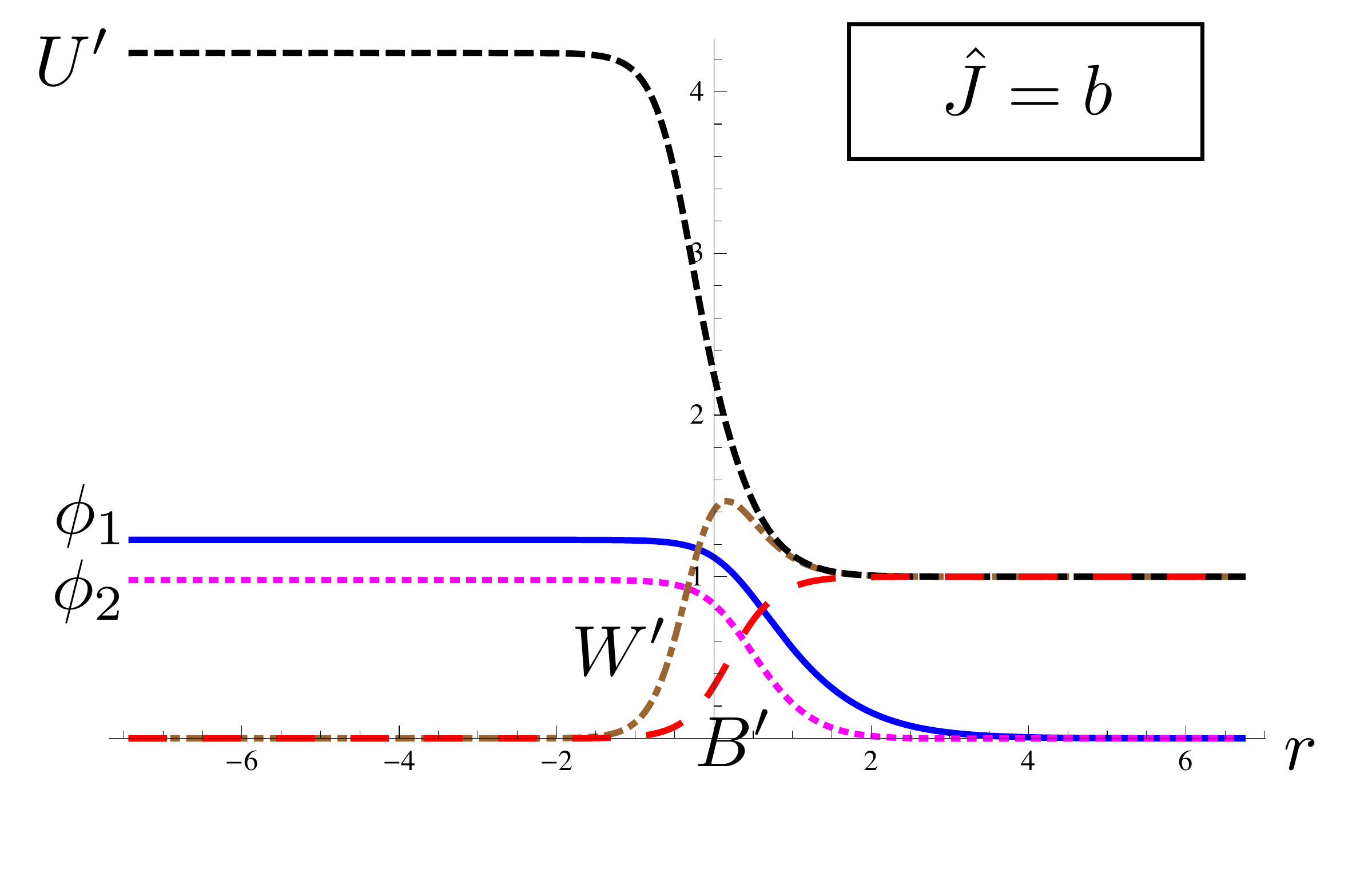} \qquad \includegraphics[width=0.45\textwidth]{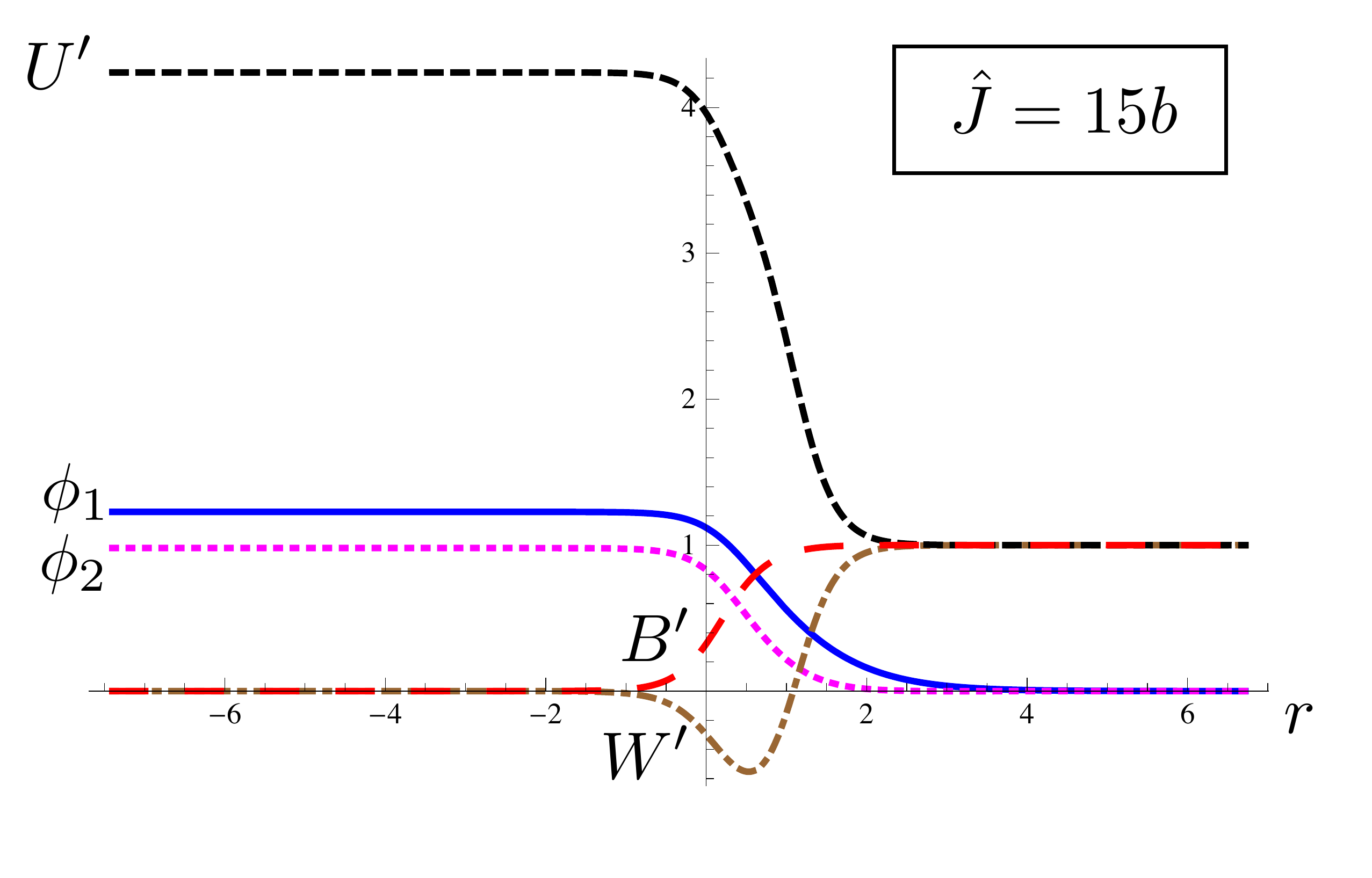} \caption{The case of two scalars witht non-vanishing $\hat J$.\label{2scalarsJ}}
\end{figure}
Finally, let us look at the more general case, where we have two running scalars and a constant non-vanishing $\hat J$. We choose the same values for the magnetic fields as in the last example, i.e.\ \eqref{P1P2P3}. Given that \eqref{P-g-X} does not depend on $\hat J$ at all, it is not surprising that this leads to the same values for the $X^A_{(0)}$ as in \eqref{XA0s} (and, thus, also \eqref{phi1phi2alpha} does not change). The main change arises for $U, W$ and $C$, as their flow equations explicitly depend on $\hat J$. They take on the near-horizon form 
\bea \label{U0W0C0J}
&& U_{(0)} = -\frac12 \ln \left(\frac{\hat J}{\alpha} + \frac{4 e^{\alpha r}}{\alpha^3} \right) + \alpha r \quad , \quad W_{(0)} = \frac12 \ln \left(\frac{\hat J}{\alpha} + \frac{4 e^{\alpha r}}{\alpha^3} \right) \ ,\nonumber \\
&& C_{(0)} = \frac{\alpha^3}{4} \left( \frac{1}{\hat{J} \frac{\alpha^2}{4} e^{-\alpha r} + 1} - 1 \right). 
\eea
Notice, in particular, the different behavior of $U_{(0)}$ and $W_{(0)}$ very close to the horizon, i.e.\ for $r \rightarrow -\infty$. Whereas the slope of $U_{(0)}$ and $W_{(0)}$ was $\alpha/2$ in the case of vanishing $\hat J$, cf.\ \eqref{U0W0C0}, now it is $\alpha$ for $U_{(0)}$ and zero for $W_{(0)}$ in the case of non-vanishing $\hat J$. We will clearly see this in the numerical solutions. 

Again, we perturb around the near-horizon solution by \eqref{perturbansatz}. We again infer that $c_2 = c_4$ and that $c_1$ and $c_3$ are related as in \eqref{perturb2}. Moreover, $\delta U, \delta W$ and $\delta B$ are all proportional to $e^{c_2 r}$. Without going into the details, we present the resulting numerical solutions for different values of $\hat J$ in figure \ref{2scalarsJ}. All these plots were produced using $c_3 = -0.1$ and $b = 4 \alpha^{-3}$. One can nicely see that the main difference to the case of vanishing $\hat J$ appears in the $U$ and $W$ sector. The different slope of $U$ and $W$ close to the horizon, mentioned in the last paragraph, is apparent. It is also obvious that for small $\hat J$, $U$ and $W$ first behave as in the case with vanishing $\hat J$ when approaching the horizon from infinity. I.e.\ they start out showing the same slope of $\alpha/2$ until the $\hat J$-term starts dominating very close to the horizon, where the slope of $U$ doubles and $W$ becomes constant. With increasing $\hat J$ the intermediate region, where $U$ and $W$ have the same slope of $\alpha/2$, becomes smaller and smaller and finally disappears altogether. 

\begin{figure}[t]
\centering \includegraphics[width=0.5\textwidth]{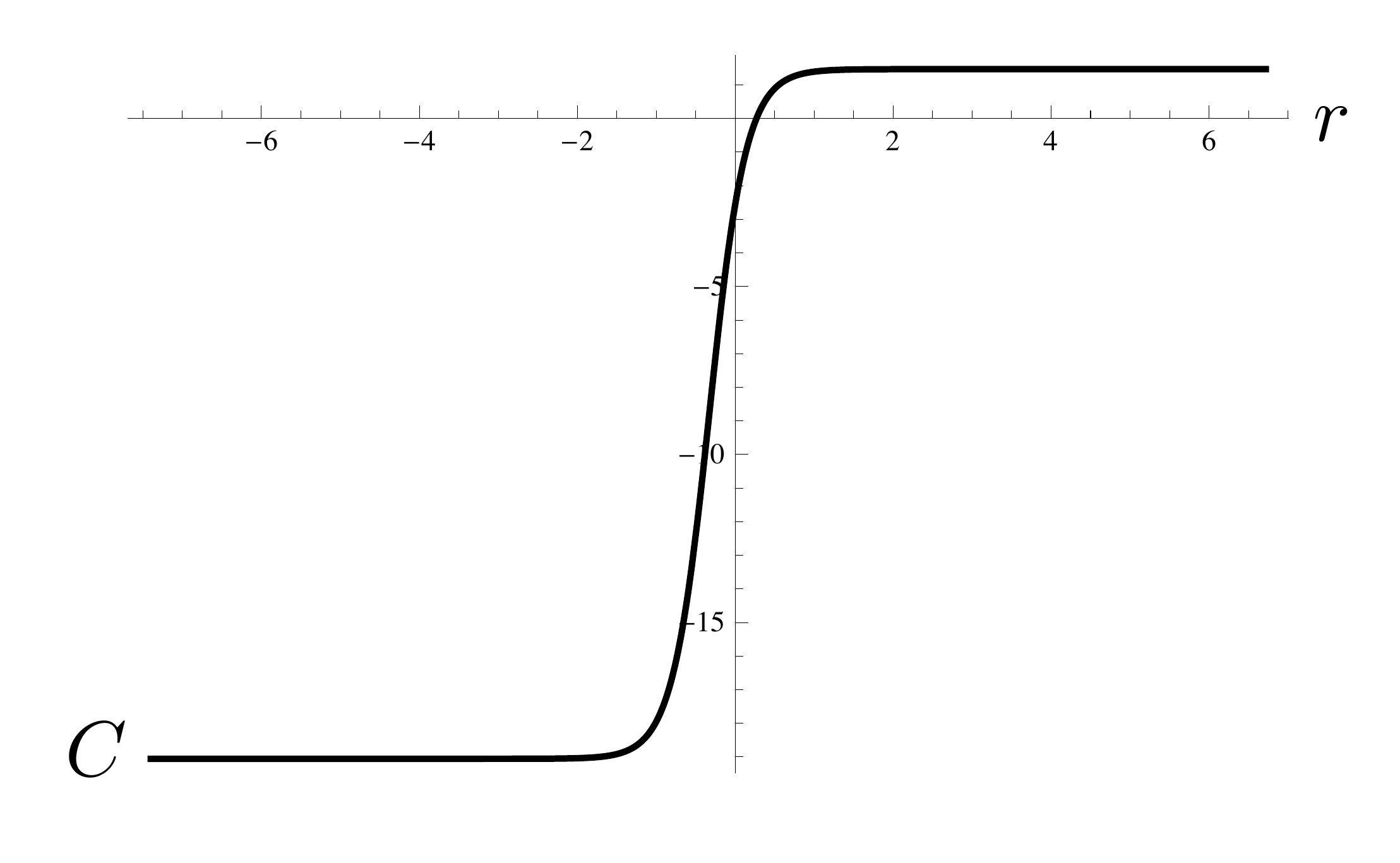}  \caption{The function $C$ for $\hat J = b$.\label{2scalars_C_Jeqb}}
\end{figure}

The function $C = e^{U-W}-1/b$ is shown (for $\hat J = b$) in figure \ref{2scalars_C_Jeqb}. Obviously, asymptotically it becomes constant and, thus, the asymptotic region is indeed given by $AdS_5$.

\begin{figure}[t]
\centering \includegraphics[width=0.45\textwidth]{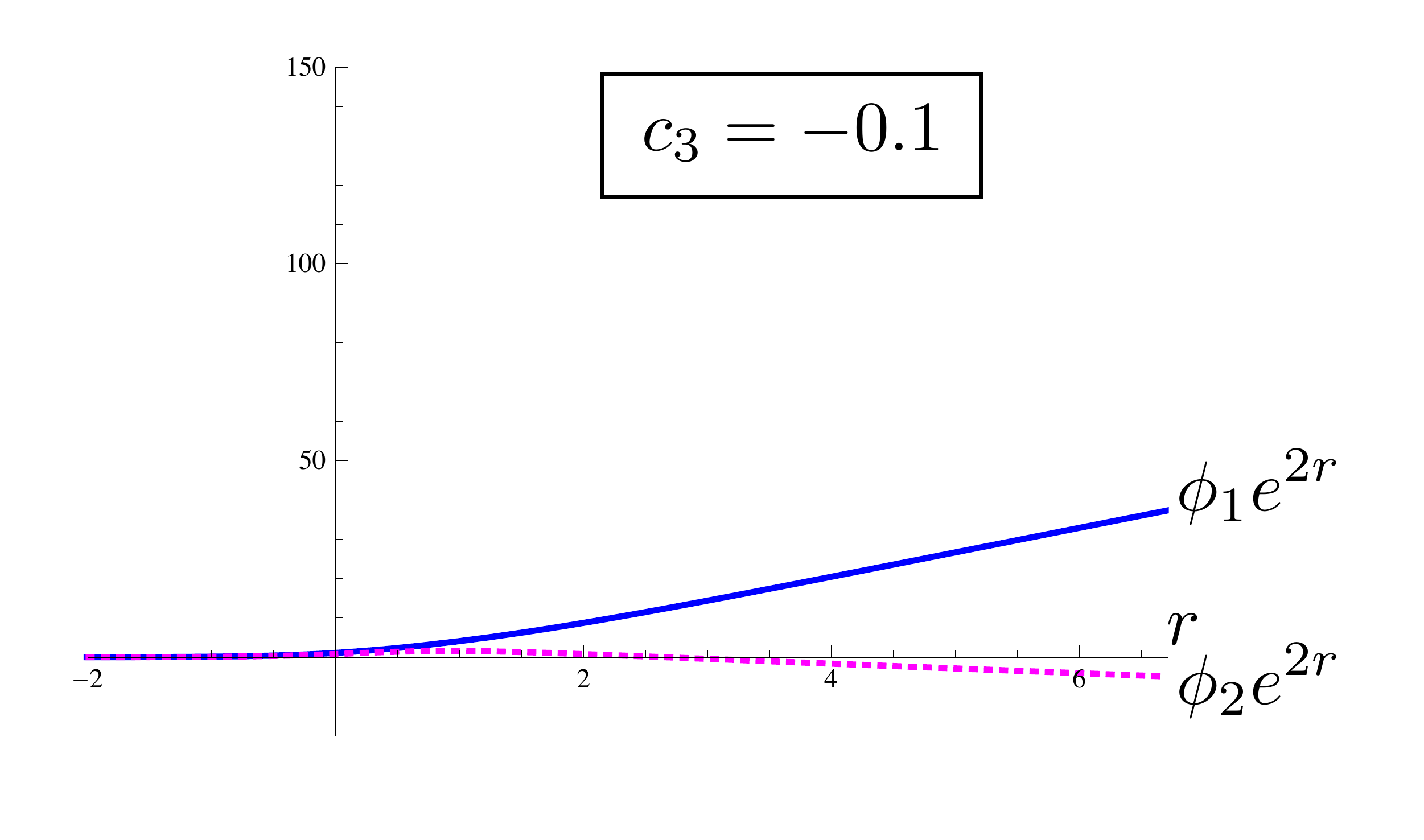}  \qquad \includegraphics[width=0.45\textwidth]{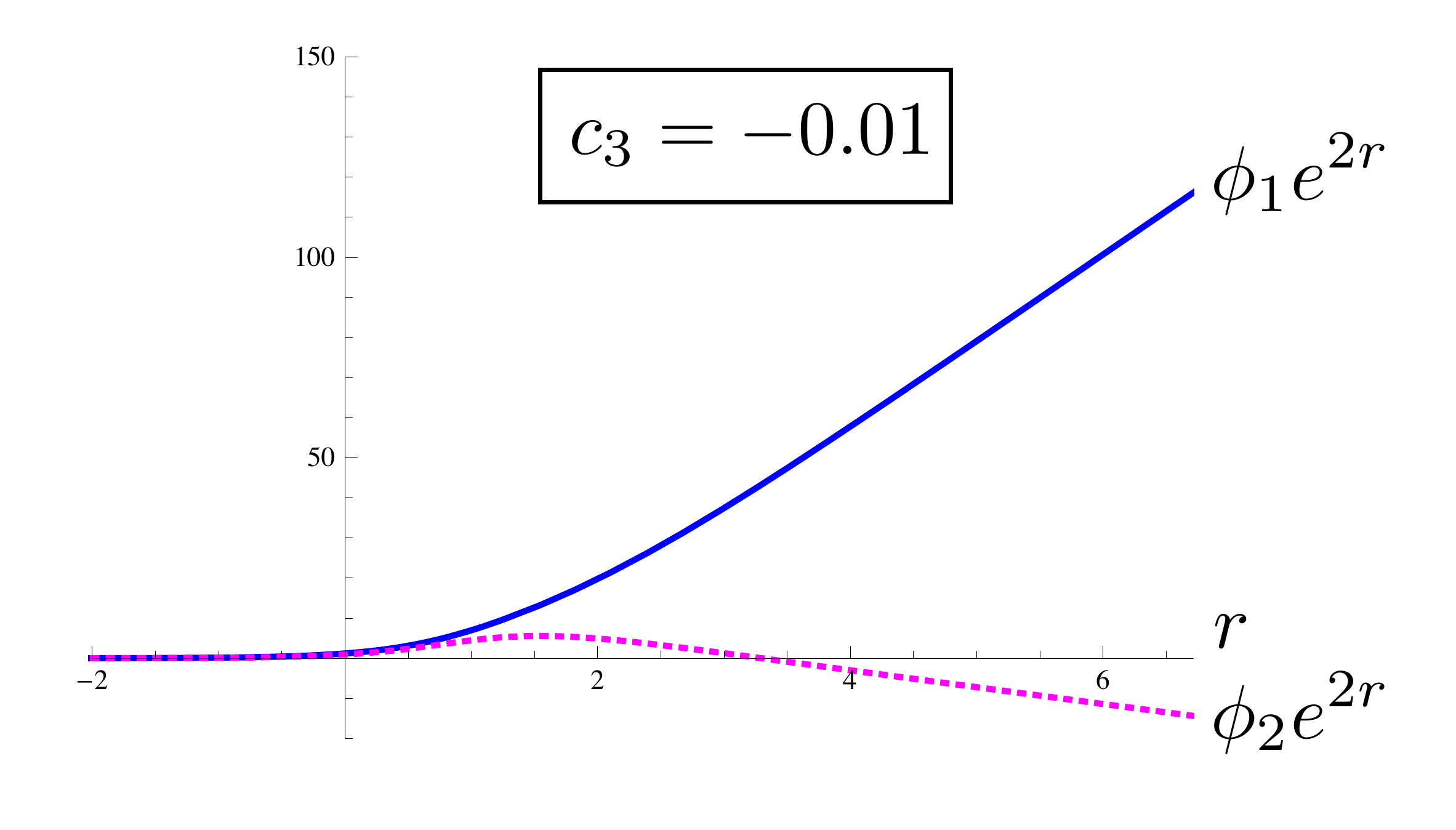} \caption{The two scalars multiplied with $e^{2 r}$.\label{2scalarsJ_asymptotia}}
\end{figure}

Finally, in figure \ref{2scalarsJ_asymptotia}, we plot $\phi^1$ and $\phi^2$, multiplied with $e^{2r}$, for two different values of $c_3$. As expected from  \eqref{sourcevev}, the graphs show a linear behavior with non-vanishing sources and 1-point functions for the two operators dual to the scalars. Obviously, these sources and 1-point functions depend on the value of $c_3$. 

\vskip 5mm

\subsection*{Acknowledgements}
We acknowledge helpful discussions with Costas Bachas, Gianguido Dall'Agata, Oscar Dias, Pau Figueras, Vishnu Jejjala,  Amir-Kian Kashani-Poor, Niels Obers, Giuseppe Policastro, Harvey Reall and Jan Troost. 
The work of G.L.C.\ was
partially supported by the Center for Mathematical Analysis, Geometry
and Dynamical Systems (IST/Portugal), as well as 
by Funda\c{c}\~{a}o para a Ci\^{e}ncia e a Tecnologia
(FCT/Portugal) through grants CERN/FP/116386/2010 and PTDC/MAT/119689/2010.  
The work of S.B., M.H.\ and S.N.\ was supported by the Excellence Cluster ``The Origin and the Structure of the Universe'' in Munich.
The work of M.H.\ and S.N.\ was also supported by the German Research Foundation (DFG) within the Emmy-Noether-Program (grant number: HA 3448/3-1). The work of S.B., G.L.C., M.H.\ and S.N.\
was supported in part by the transnational cooperation FCT/DAAD grant ``Black Holes, duality and string theory".


\begin{appendix}

\section{Einstein equations \label{sec:ricci}}

When evaluated on the solution ansatz \eqref{metricrotation}, 
the independent Einstein equations take the following form:
\par\noindent $tt$-component:
\begin{eqnarray}\label{einsteintt}
\lefteqn{\frac12 e^{-2 (U + V)} \bigg(-e^{2 (U + W)} C'^2 - 2 e^{2 (U + W)} C (2 B'C'-C'(U'+V'-3W')+C'')}\nonumber\\
&& + 2 e^{4 U} (2 B'U'+U'^2-U'V'+U'W'+U'')\nonumber\\
&& - e^{2 W} C^2 \big(e^{2 W} C'^2 + 2 e^{2 U} (2 B'W'+U'W'-V'W'+W'^2+W'')\big)\bigg)\nonumber\\
&=&(-e^{2U}+e^{2W}C^2)\bigg(\frac13g^2\left(G^{AB}h_Ah_B-2(h_AX^A)^2\right)-\frac16 G_{AB}F^A_{KL}F^{B\,KL}\bigg)+G_{AB}F^A_{rt}F^B_{rt}e^{-2V}\nonumber\\
&&
\end{eqnarray}
$rr$-component:
\begin{eqnarray}\label{einsteinrr}
\lefteqn{-2 B'^2+\frac12 e^{2W-2U}C'^2-U'^2+2B'V'+U'V'+V'W'-W'^2-2 B''-U''-W''}\nonumber\\
&=&G_{AB}(X^A)'(X^B)'+e^{2V}\left(\frac13g^2\left(G^{AB}h_Ah_B-2(h_AX^A)^2\right)-\frac16 G_{AB}F^A_{KL}F^{B\,KL}\right)\nonumber\\
&&+G_{AB}\left(F^A_{rt}F^B_{rt}(-e^{-2U})+(A^A_z)'(A^B_z)'(e^{-2W}-e^{-2U}C^2)+2F^A_{rt}(A^B_z)'e^{-2U}C\right)\nonumber\\
&&
\end{eqnarray}
$xx$-component:
\begin{eqnarray}\label{einsteinxx}
&& -e^{2B-2V}\left(2B'^2+B'U'-B'V'+B'W'+B''\right)= \nonumber\\
&&e^{2B}\left(\frac13g^2\left(G^{AB}h_Ah_B-2(h_AX^A)^2\right)-\frac16 G_{AB}F^A_{KL}F^{B\,KL}\right)
+G_{AB}P^AP^Be^{-2B}
\end{eqnarray}
$zz$-component:
\begin{eqnarray}\label{einsteinzz}
\lefteqn{\frac12 e^{2W-2U-2V}\left(-e^{2W}C'^2-2e^{2U}(2B'W'+U'W'-W'V'+W'^2+W'')\right)}\nonumber\\
&=&e^{2W}\left(\frac13g^2\left(G^{AB}h_Ah_B-2(h_AX^A)^2\right)-\frac16 G_{AB} F^A_{KL}F^{B\,KL}\right)+G_{AB}(A^A_z)'(A^B_z)'e^{-2V}\nonumber\\
&&
\end{eqnarray}
$tz$-component:
\begin{eqnarray}\label{einsteintz}
\lefteqn{\frac12 e^{2W-2U-2V}\bigg(-2e^{2U}(B'C'+2CB'W')+e^{2U}(C'U'+C'V'-3C'W'-C'')-e^{2W}CC'^2}\nonumber\\
&&+2e^{2U}C(U'W'-V'W'+W'^2+W'')\bigg)\nonumber\\
&=&e^{2W}C\left(\frac13g^2\left(G^{AB}h_Ah_B-2(h_AX^A)^2\right)-\frac16 G_{AB}F^A_{KL}F^{B\,KL}\right)+2G_{AB}F^A_{rt}(A^B_z)'e^{-2V}\nonumber\\
&&
\end{eqnarray}


\section{Relating five- and four-dimensional flow equations \label{45dict}}

We relate the four-dimensional flow equations for black branes derived in \cite{Barisch:2011ui} in big moduli space to the 
five-dimensional flow equations \eqref{eqndyonic}.  We set $g=1$ throughout.

The four-dimensional $N = 2$, $U(1)$ gauged supergravity theory is based on complex scalar fields $X^I$ encoded in
the 
cubic prepotential (with $A = 1, \dots, n$)
\begin{equation}
F(X)=-\frac16 \frac{C_{ABC}X^AX^BX^C}{X^0}=(X^0)^2{\cal F}(z) \;,
\end{equation}
where $z^A = X^A/X^0$ denote the physical scalar fields and
\begin{equation}
{\cal F}(z)=-\frac16 C_{ABC}z^Az^Bz^C\,.
\end{equation}
Differentiating with respect to $z^A$ yields
\begin{eqnarray}
{\cal F}_A&=&-\frac12C_{ABC}z^Bz^C\,,\nonumber\\
{\cal F}_{AB}&=&-C_{ABC}z^C\,,
\end{eqnarray}
where ${\cal F}_A = \partial {\cal F}/\partial z^A$, etc.  The K\"ahler potential $K(z, \bar z)$ is determined in terms of 
${\cal F}$ by 
\begin{eqnarray}
e^{-K}&=& i \left(2({\cal F}-\bar{{\cal F}})-(z^A-\bar{z}^A)({\cal F}_A+\bar{{\cal F}}_A)\right)\nonumber\\
&=&\frac{i}{6}C_{ABC}(z^A-\bar{z}^A)(z^B-\bar{z}^B)(z^C-\bar{z}^C)\,.
\end{eqnarray}
The K\"ahler metric $g_{A\bar{B}} = \partial_A \partial_{\bar B} K(z, \bar z)$ can be expressed as
\begin{eqnarray}
g_{A\bar{B}}&=&K_AK_{\bar{B}}-ie^{K}\left({\cal F}_{AB}-\bar{{\cal F}}_{AB}\right)\;,
\end{eqnarray}
where $K_A = \partial K / \partial z^A$ and
\begin{equation}
K_A = - K_{\bar A} = - e^{K} \, \tfrac{i}{2} \, C_{ABC} (z^B - {\bar z}^B) (z^C - {\bar z}^C) \;.
\end{equation}
In the following we pick the gauge $X^0(z) =1, X^A(z) = z^A$ (with $X^I(z) \equiv X^I e^{-K/2}$), so that the complex scalar fields $X^I$ and the $z^A$ are related by $X^0 = e^{K/2}$ and $X^A = e^{K/2} \, z^A$.

Using the dictionary \eqref{4d-5d-rel} that relates the quantities appearing in the four- and five-dimensional 
flow equations, in particular $z^A - \bar z^A = 2 i e^W \, X^A_{\rm 5d}$, we obtain
\begin{equation}
e^{-K} = 8 v \, e^{3W}
\label{K-W}
\end{equation}
as well as 
\begin{eqnarray}
K_{\bar A} &=& - \frac{3i}{2v} \, e^{-W} \, X_A^{\rm 5d} \;, \nonumber\\ 
g_{A\bar{B}} &=& 4ve^{K + W}G_{AB}=\frac12e^{- 2 W}G_{AB}\,,
\label{KA-gABB}
\end{eqnarray}
where $G_{AB}$ denotes the target space metric in five dimensions, cf.\ \eqref{GAB-v}, and the 
$X_A^{\rm 5d}$ were defined in \eqref{XA-v}. The factor of $v$ in \eqref{K-W} arises due to the normalization in \eqref{vnorm}.
Using these expressions, we establish
\begin{equation}
g^{A \bar B} \, K_{\bar B} = - (z^A - {\bar z}^A) \;.
\label{special-g-G}
\end{equation}

In the big moduli space, the four-dimensional flow equations were expressed in terms of rescaled
complex scalar fields $Y^I$ given by $Y^0 = |Y^0| e^{i \alpha}$ and $Y^A = Y^0 \, z^A$, where
$|Y^0| = e^{K/2 + \psi - U_4}$.  On a solution to the four-dimensional flow equations we
can relate the phase $\alpha$ to the phase $\gamma$ that enters in the four-dimensional flow
equations.  We obtain $\alpha = - \gamma$, which we establish as follows. Writing
$e^{2 i \alpha} = Y^0/\bar{Y}^0$ we get
\bea \label{alphaprime}
\alpha' = - \frac{i}{2} e^{-2 i \alpha} \left( \frac{(Y^0)'}{\bar{Y}^0} - (\bar{Y}^0)' \frac{Y^0}{(\bar{Y}^0)^2}  \right) = - \frac{i}{2 \bar{Y}^0} \left( e^{-2 i \alpha} (Y^0)' - (\bar{Y}^0)' \right) .
\eea
The flow equation for $Y^0$ reads \cite{Barisch:2011ui}
\begin{eqnarray}
(Y^0)' &=& e^{\psi - U_4} \, N^{0J} \, {\bar q}_J \nonumber\\
&=& e^{\psi- U_4 + K} \left[ g^{A \bar B} \, K_A \, \bar{q}_{\bar J} \, ( \partial_{\bar B} + K_{\bar B} ) \, {\bar X}^{\bar J} 
(\bar{z}) - \bar{q}_{\bar J} \, {\bar X}^{\bar J}(\bar{z}) \right] \;,
\label{eq:flow-Y}
\end{eqnarray}
where we used
\bea \label{NIJBernard}
N^{IJ} = e^K \Big[ g^{A \bar B}\, (\partial_A + \partial_A K) X^I(z)\, (\partial_{\bar B} + \partial_{\bar B} K) \bar X^J(\bar z) - X^I(z) \bar X^J(\bar z) \Big]\ ,
\eea
cf.\ for instance \cite{deWit:1996ag}. In \eqref{eq:flow-Y} the $q_I$ denote the four-dimensional quantities
\begin{equation}
q_I = e^{U_4 - 2 \psi + i \gamma} \left(\hat{Q}_I - i e^{2(\psi - U_4)} \, \hat{h}_I \right) \;,
\label{q-4d-hat}
\end{equation}
which should not be confused with the five-dimensional electric charges $q_A^{\rm 5d}$. 
The quantities $\hat Q_I$ and $\hat h_I$ are combinations of the 
four-dimensional charges and fluxes given by  \cite{Barisch:2011ui}
\begin{eqnarray}
\hat{Q}_I &=& Q_I - F_{IJ} \, P^J \;, \nonumber\\
\hat{h}_I &=& h_I - F_{IJ} \, h^J \;.
\end{eqnarray}
For later use, we also introduce the quantities
\begin{eqnarray}
Z(Y) &=& - \hat{Q}_I \, Y^I \;, \nonumber\\
W(Y) &=& - \hat{h}_I \, Y^I \;.
\label{Z-W}
\end{eqnarray}
Inserting the flow
equation \eqref{eq:flow-Y} in \eqref{alphaprime} yields
\bea \label{alphaprime2}
\alpha' &=& - \frac{i}{2} e^{i \alpha} e^{K/2} \left[ \left(g^{A \bar{B}} K_A K_{\bar{B}} -1\right) \left( e^{-2 i \alpha} (\bar{q}_0 + \bar{q}_{\bar{C}} \bar{z}^{\bar{C}}) - (q_0 + q_{C} z^{C}) \right) \right. \nonumber \\
&& \hspace{2cm} \left. + g^{A \bar{B}} ( e^{-2 i \alpha} K_A \bar{q}_{\bar{B}} -  q_A K_{\bar{B}}) \right]\ ,
\eea
where we used the relation
\begin{equation}
\bar{q}_J\left(\partial_{\bar{B}}+K_{\bar{B}}\right)\bar{X}^J(z)=\bar{q}_{\bar{B}}+K_{\bar{B}}\left(\bar{q}_0+\bar{q}_A\bar{z}^A\right)
\label{pderKq}
\end{equation}
as well as
$|Y^0| = e^{K/2 + \psi - U_4}$.  Next, using that on a four-dimensional solution we have $q_I Y^I = {\bar q}_I {\bar Y}^{\bar I}$, we obtain
\begin{eqnarray}
e^{- 2 i \alpha} \, ({\bar q}_0 + {\bar q}_{\bar C} {\bar z}^{\bar C}) =
q_0 + q_C z^C\ .
\label{eq:qbq-rel}
\end{eqnarray}
Inserting this in \eqref{alphaprime2} results in
\bea \label{alphaprime3}
\alpha' = - \frac{i}{2} e^{i \alpha} e^{K/2} g^{A \bar{B}} \left( e^{-2 i \alpha} K_A \bar{q}_{\bar{B}} -  q_A K_{\bar{B}}\right) \ .
\eea
Using \eqref{special-g-G}, we obtain
\begin{eqnarray}
 g^{A \bar B} \left( e^{- 2 i \alpha} \,  K_A \, {\bar q}_{\bar B} - q_A \, K_{\bar B} \right)&=& 
 (z^A - {\bar z}^A) \left(e^{- 2 i \alpha} \, {\bar q}_{\bar A} + q_A \right) \nonumber \\
 & = & e^{- 2 i \alpha} z^A {\bar q}_{\bar A} - {\bar z}^A q_A  - e^{- 2 i \alpha}  {\bar z}^A  {\bar q}_{\bar A} + q_A z^A \nonumber \\
 & = & e^{- 2 i \alpha} ({\bar q}_0 + z^A {\bar q}_{\bar A})- (q_0 + {\bar z}^A q_A)\ ,
 \end{eqnarray}
where we used \eqref{eq:qbq-rel} in the last equality. Now we notice that 
\bea
q_I \bar{z}^I  = q_0 + {\bar z}^A q_A = - e^{i(\alpha+\gamma)} e^{-K/2+2U_4-3\psi} \left(\bar{Z} (\bar{Y}) - i 
e^{2 (\psi - U_4)} \bar{W}(\bar{Y})\right)
\eea
and 
\bea
e^{- 2 i \alpha} \bar{q}_{\bar{I}} z^I  = e^{- 2 i \alpha} (\bar{q}_0 + z^A \bar{q}_{\bar{A}}) = - e^{-i(3 \alpha+\gamma)} e^{-K/2+2U_4-3 \psi} \left(Z (Y) + i e^{2 (\psi - U_4)} W(Y)\right) ,
\eea
so that
\bea  \label{alphaprime4}
\alpha' = - e^{2U_4 - 3 \psi}\, {\rm Im} \left(e^{- i (2 \alpha + \gamma)} Z(Y)\right) - e^{-\psi} \, {\rm Re} \left(e^{- i (2 \alpha + \gamma)} W(Y) \right)\ .
\eea
This is precisely the flow equation for $\gamma$, provided $\alpha = - \gamma$.

Using this result, we now relate the flow equations for the $z^A$ to the five-dimensional flow equations for $X^A_{\rm 5d}$
and $A_z^A$. Using the four-dimensional flow equations for $Y^0$ and $Y^A$ we obtain
\begin{eqnarray}
(z^A)'&=&\frac{1}{Y^0}\left((Y^A)'-z^A(Y^0)'\right)\nonumber\\
&=&\frac{e^{\psi-U_4}}{Y^0}\left(N^{AJ}-z^AN^{0J}\right)\bar{q}_J\nonumber\\
&=&e^{-\frac{K}{2}+i\gamma}\left(N^{AJ}-z^AN^{0J}\right)\bar{q}_J\,.
\end{eqnarray}
Then, using \eqref{NIJBernard}, one derives
\begin{equation}
N^{AJ}-z^AN^{0J}=e^Kg^{A\bar{B}}\left(\partial_{\bar{B}}+K_{\bar{B}}\right)\bar{X}^J(z) \;,
\end{equation}
which implies
\begin{equation}\label{eqzA}
(z^A)'=e^{\frac{K}{2}+i\gamma}g^{A\bar{B}}\bar{q}_J\left(\partial_{\bar{B}}+K_{\bar{B}}\right)\bar{X}^J(z) \;.
\end{equation}
Now we specialize to four-dimensional solutions that are supported by electric charges $Q_I$, magnetic charges $P^A$
and electric fluxes $h_A$.  Decomposing $z^A = C^A + i \hat{X}^A$ and 
using the expression \eqref{q-4d-hat} gives
\begin{eqnarray}
&& \bar{q}_0+\bar{q}_A\bar{z}^A =e^{U_4-2\psi-i\gamma}\bigg[Q_0+\frac12 C_{ABC}P^AC^BC^C+Q_AC^A-\frac12 C_{ABC} P^A\hat{X}^B\hat{X}^C \nonumber\\
&& \qquad +e^{2(\psi-U_4)}h_A\hat{X}^A
+i\left(-Q_A\hat{X}^A-C_{ABC}P^AC^B\hat{X}^C+e^{2(\psi-U_4)}h_AC^A\right)\bigg] \;.
\end{eqnarray}
Then, using \eqref{pderKq} and \eqref{KA-gABB} leads to (we recall \eqref{phi-W-rel})
\begin{eqnarray}
(z^A)'&=&2e^{\frac{K}{2}+U_4-2\psi-4\phi}G^{AB}\bigg[-12ie^{K-4\phi}X_{B}^{\rm 5d}\bigg(Q_0+\frac12 C_{CDE}P^CC^DC^E\nonumber\\
&&\phantom{2e^{\frac{K}{2}+U_4-2\psi-4\phi}G^{AB}\bigg[}+Q_EC^E-\frac12 C_{CDE}P^CX^D_{\rm 5d}X^E_{\rm 5d}e^{-4\phi}+e^{2(\psi - U_4)}h_EX^E_{\rm 5d} \, e^{-2\phi}\bigg)\nonumber\\
&&\phantom{2e^{\frac{K}{2}+U_4-2\psi-4\phi}G^{AB}\bigg[}+Q_B+C_{BEF}C^EP^F\nonumber\\
&&
+12e^{K-4\phi}X_{B}^{\rm 5d}\bigg(-Q_EX^E_{\rm 5d}e^{-2\phi}-C_{CDE}P^CC^DX^E_{\rm 5d}e^{-2\phi}+e^{2(\psi-U_4)}h_EC^E
\bigg)\nonumber\\
&&\phantom{2e^{\frac{K}{2}+U_4-2\psi-4\phi}G^{AB}\bigg[}+i\left(e^{2(\psi-U_4)}h_B-C_{BEF}P^EX^F_{\rm 5d}e^{-2\phi}\right)\bigg]\,.
\end{eqnarray}
Thus, we obtain for the real part,
\begin{eqnarray}\label{eqimaginary}
(C^A)'&=&2e^{\frac{K}{2}+U_4-2\psi-4\phi}G^{AB}\bigg[Q_B+C_{BEF}C^EP^F\nonumber\\
&&
+12e^{K-4\phi}X_{B}^{\rm 5d}\bigg(-Q_E\hat{X}^E-C_{CDE}P^CC^D\hat{X}^E+e^{2(\psi-U_4)}h_EC^E\bigg)\bigg]\,.\nonumber\\
&&
\end{eqnarray}
Next we show that the second line of this equation vanishes by virtue of the four-dimensional flow constraint
\begin{equation}
{\rm Im}\left(e^{i\gamma}Z(Y)\right)-e^{2(\psi-U_4)}{\rm Re}\left(e^{i\gamma}W(Y)\right)=0\, .
\label{constraingZ-W}
\end{equation}
We have
\begin{equation}
Z(Y)=Y^0\left(-\frac12C_{ABC}P^Az^Bz^C-Q_Az^A - Q_0 \right)
\end{equation}
and 
\begin{equation}
W(Y)=-Y^0h_Az^A\,.
\end{equation}
This leads to
\begin{equation}
{\rm Im}\left(e^{i\gamma}Z(Y)\right)=|Y^0|\left(-C_{ABC}P^AC^B\hat{X}^C-Q_A\hat{X}^A\right)
\end{equation}
as well as
\begin{equation}
{\rm Re}\left(e^{i\gamma} \, W(Y) \right)=-|Y^0|h_AC^A\,.
\end{equation}
This gives 
\begin{equation}|Y^0|\left(-C_{ABC}P^AC^B\hat{X}^C-Q_A\hat{X}^A+e^{2(\psi-U_4)}h_AC^A\right)=\nonumber\end{equation}
\begin{equation}
={\rm Im}\left(e^{i\gamma}Z(Y)\right)-e^{2(\psi-U_4)}{\rm Re}\left(e^{i\gamma}W(Y)\right) \;,
\end{equation}
which vanishes due to \eqref{constraingZ-W}, so that \eqref{eqimaginary} becomes
\begin{eqnarray}
(C^A)' =\frac{1}{\sqrt{2v}}e^{V-2B}G^{AB}\left(Q_B+C_{BEF}C^EP^F\right) \;,
\label{flow-C4d}
\end{eqnarray}
where we used the relations \eqref{rel-warp} and \eqref{K-W}.

For the imaginary part of $(z^A)'$ we get
\begin{eqnarray}\label{realpart1}
\left(e^{-2\phi}X^A_{\rm 5d}\right)'&=&2e^{\frac{K}{2}+U_4-2\psi-4\phi}\bigg[-e^{2\phi}X^A_{\rm 5d}\bigg(Q_0+\frac12 C_{BCD}P^BC^CC^D+Q_BC^B\nonumber\\
&&\phantom{2e^{\frac{K}{2}+U_4-2\psi-4\phi}\bigg[}-\frac12C_{BCD}P^BX^C_{\rm 5d}X^D_{\rm 5d}e^{-4\phi}+e^{2(\psi - U_4)}h_BX^B_{\rm 5d}e^{-2\phi}\bigg)\nonumber\\
&&\phantom{2e^{\frac{K}{2}+U_4-2\psi-4\phi}\bigg[}+G^{AB}\left(e^{2(\psi - U_4)}h_B-C_{BEF}P^EX^F_{\rm 5d}e^{-2\phi}\right)\bigg]\,.
\end{eqnarray}
Using \eqref{GAB-v} we obtain
\begin{equation}
-C_{ABC}X^C_{\rm 5d}=2vG_{AB}-\frac{9}{v}X_{A}^{\rm 5d}X_{B}^{\rm 5d}\,.
\end{equation}
Contracting this expression once with $P^AX^B_{\rm 5d}$ and once with $P^A$, we rewrite the two expressions containing $C_{BCD}P^BX^C_{\rm 5d}X^D_{\rm 5d}$ and $C_{BEF}P^EX^F_{\rm 5d}$ in \eqref{realpart1}. Using 
\eqref{K-W} as well we obtain
\begin{eqnarray}\label{realpart2}
\left(e^{-2\phi}X^A_{\rm 5d}\right)'&=&\frac{1}{\sqrt{2v}}e^{U_4-2\psi-\phi}\bigg[-e^{2\phi}X^A_{\rm 5d}\bigg(Q_0+\frac12 C_{BCD}P^BC^CC^D+Q_BC^B\bigg)\nonumber\\
&&\phantom{2e^{\frac{K}{2}+U_4-2\psi-4\phi}\bigg[}+2vP^Ae^{-2\phi}-3X^A_{\rm 5d}(X_B^{\rm 5d}P^B)e^{-2\phi}\nonumber\\
&&\phantom{2e^{\frac{K}{2}+U_4-2\psi-4\phi}\bigg[}+e^{2(\psi-U_4)}G^{AB}\left(h_B-\frac{3}{2v}X_{B}^{\rm 5d}(h_CX^C_{\rm 5d})\right)\bigg]\nonumber\\
&=&\left(e^{-2\phi}\right)'X^A_{\rm 5d}+e^{-2\phi}(X^A_{\rm 5d})'\,.
\end{eqnarray}
Using $X_{A}^{\rm 5d}(X^A_{\rm 5d})'=0$ (and \eqref{phi-W-rel}, \eqref{rel-warp}) we infer
\begin{eqnarray}\label{matcheqW}
\left(e^{-2\phi}\right)' &=&\frac{1}{\sqrt{2v}}e^{U_4-2\psi-\phi}\bigg[-e^{2\phi}\left(Q_0+\frac12 C_{BCD}P^BC^CC^D+Q_BC^B\right)\nonumber\\
&&\phantom{\frac{1}{\sqrt{2v}}e^{U_4-2\psi-\phi}X^A_{5d}\bigg[}-e^{-2\phi}X_{B}^{\rm 5d}P^B-\frac{1}{3}e^{2(\psi-U_4)}h_CX^C_{\rm 5d}\bigg]\nonumber\\
&=&-\frac{1}{\sqrt{2v}}\bigg[e^{-2B-2W+V-2\phi}\left(Q_0+\frac12 C_{BCD}P^BC^CC^D+Q_BC^B\right)\nonumber\\
&&\phantom{-\frac{1}{\sqrt{2v}}X^A_{\rm 5d}\bigg[}+e^{V-2B-2\phi}X_{B}^{\rm 5d}P^B+\frac13e^{-U}h_CX^C_{\rm 5d}\bigg]
\end{eqnarray}
as well as 
\begin{eqnarray}
(X^A_{\rm 5d})'&=&\frac{1}{\sqrt{2v}}e^{U_4-2\psi+\phi}\bigg[2e^{-2\phi}\left(vP^A-X^A_{\rm 5d}(X_B^{\rm 5d} 
P^B)\right)\nonumber\\
&&\phantom{\frac{1}{\sqrt{2v}}e^{U_4-2\psi+\phi}\bigg[}+e^{2(\psi-U_4)}G^{AB}\left(h_B-\frac{1}{v}X_{B}^{\rm 5d}(h_CX^C_{\rm 5d})\right)\bigg]\,.
\label{flow-X5d}
\end{eqnarray}
The former should match the flow equation for $e^W$. To check this, we note that the five-dimensional
flow equations \eqref{eqndyonic} imply
\begin{eqnarray}
U'+W'=\frac23h_AX^A_{\rm 5d} e^V+\frac{1}{v}X_{A}^{\rm 5d}P^Ae^{V-2B}\,.
\end{eqnarray}
Subtracting this from the third equation of \eqref{eqndyonic} gives
\begin{equation}
W'=\frac{1}{3}h_AX^A_{\rm 5d}e^V+\frac{1}{2v}X_{A}^{\rm 5d}P^Ae^{V-2B}-\frac{1}{2}\hat{J}e^{-2B-2W+V} \;,
\end{equation}
so that
\begin{equation}
\left(e^W\right)'=e^{-2\phi}W'=\frac13h_AX^A_{\rm 5d}e^{V-2\phi}+\frac{1}{2v}X_{A}^{\rm 5d}P^Ae^{V-2B-2\phi}-\frac12\hat{J}e^{-2B-2W+V-2\phi}\,.
\end{equation}
This matches \eqref{matcheqW} if we set $v=1/2$ and perform the identifications \eqref{4d-5d-rel}.
Under these identifications, the flow equations \eqref{flow-C4d} and \eqref{flow-X5d} precisely match those 
for $A_z^A$ and $X^A_{\rm 5d}$ appearing in 
\eqref{eqndyonic}.  Similarly, the flow equations for the four-dimensional warp factors $U_4$ and $\psi$ match
those of the five-dimensional warp factors $U$ and $B$ using \eqref{rel-warp}.


\section{A different first-order rewriting \label{electric-rew}}

We present a different first-order rewriting that allows for solutions with electric fields.
This rewriting is the one performed in \cite{Cardoso:2008gm} for static black hole solutions, which we adapt
to the case of stationary black branes in the presence of 
magnetic fields.

We consider the metric \eqref{metricrotation} and the gauge field ansatz \eqref{gaugefielddyonic} with $A^A_z=0$,
so that $\hat{q}_A = q_A$ and $\hat{J} = J$.  
The starting point of the analysis is therefore Lagrangian \eqref{lagrange2}, 
with $A_z^A=0$, $\hat{q}_A = q_A$ and $\hat{J} = J$.

We perform the following $g$-split of $U$ and $V$, 
\begin{eqnarray}
U&=&U_0+\frac12 \log f\,,\nonumber\\
V&=&V_0-\frac12 \log f\,,\nonumber\\
f &=&f_0(r)+g^2f_2(r)=-\mu \, r^2+g^2 e^{2 U_2(r)}\,, \nonumber\\
V_0 &=& 2B(r)+W(r)+U_0(r)+\log(r) \;.
\label{f0V0}
\end{eqnarray}
In addition, we perform the rescaling
\begin{equation}
P^A = g \, p^A \;\;\;,\;\;\; J = g \, j \;,
\end{equation}
and we organize the terms in the Lagrangian into powers of $g$.  This yields ${\cal L} = {\cal L}_0 + g^2 \, {\cal L}_2$.  First, we analyze ${\cal L}_0$, 
\begin{eqnarray}
{\cal L}_0&=&e^{2B+W+U_0-V_0}f_0\bigg(2B'^2+2U'W'+4B'W'+4B'U'-G_{AB}(X^A)'(X^B)'\nonumber\\
&&\phantom{e^{2B+W+U_0-V_0}f_0\bigg(}-\frac14 G^{AB} {q}_A {q}_B \, e^{-4B-2W+2V_0}(f_0)^{-1}\bigg)\,.
\end{eqnarray}
We perform a first-order rewriting of ${\cal L}_0$ by introducing parameters $\tilde{q}_A$ and  $\gamma_A$ that are related to the electric charges $q_A$ by
\begin{equation}
\frac14 G^{AB} {q}_A {q}_B=-\mu \, G^{AB} \,\tilde{q}_A\gamma_B \;.
\end{equation}
We obtain
\begin{eqnarray}
 {\cal L}_0&=&-\mu \, r \, e^{U_0} \tilde{q}_A G^{AB}\left(6 X_B+e^{U_0}(r^2\tilde{q}_B-\gamma_B)\right)\nonumber\\
&&+e^{2B+W+U_0-V_0} f_0\bigg[\frac12 (2B'+U_0') (2B'+4W'+3 U_0')\nonumber\\
&&\phantom{ndsja}-\frac94 G^{AB}\big(X_A'-U_0'X_A+\frac23 e^{-2B-W+V_0}\tilde{q}_A\big)\big(X_B'-U_0'X_B+\frac23 e^{-2B-W+V_0}\tilde{q}_B\big)\bigg]\nonumber\\
&&-2\left(e^{U_0}\tilde{q}_AX^A f_0\right)'-\mu(2W'+4B')\,.
\end{eqnarray}
This yields the first-order flow equations
\begin{eqnarray}\label{eomdyonicsplit0}
X_A'&=&U_0' X_A-\frac23 e^{-2B-W+V_0}\tilde{q}_A\,,\nonumber\\
B' &=& -\frac12 U_0'\,,\nonumber\\
2B'&=&-4W'-3 U_0'\,.
\end{eqnarray}
Using \eqref{f0V0} as well as $X^A (X_A)' =0$,  
these equations yield
\begin{eqnarray}
W' &=& B' = - \frac12 U_0' \nonumber\\
\left(e^{-U_0} \right)' &=& - \frac23 r \, X^A \tilde{q}_A \;, \nonumber\\
\left( e^{- U_0} X_A \right)' &=& - \frac23 \, r \, \tilde{q}_A \;.
\label{eq:eqs-L0}
\end{eqnarray}
Integrating the latter gives
\begin{equation}
e^{- U_0} X_A = \frac13 \, H_A \;\;\;,\;\;\; H_A = \tilde{\gamma}_A - r^2  \, \tilde{q}_A \;,
\label{solXA}
\end{equation}
where $\tilde{\gamma}_A$ denote integration constants.  Contracting this with $X^A$ results in 
\begin{equation}
e^{- U_0} = \frac13 \, H_A \, X^A \;,
\end{equation}
which satisfies the second equation of \eqref{eq:eqs-L0} by virtue of $X_A (X^A)' =0$.

${\cal L}_0$ contains, in addition, the first line, which is not
the square (or the sum of squares) of a first-order flow equation. 
 Its variation with respect to $U_0$ gives
 \begin{equation}
 \tilde{q}_A \, G^{AB} \left( 3 X_B + e^{U_0} \, \left(r^2 \, \tilde{q}_B - \gamma_B \right)\right)  = 0 \;.
 \end{equation}
Comparing with \eqref{solXA} yields
\begin{equation}
 \tilde{q}_A \, G^{AB} \, \left(\tilde{\gamma}_B - \gamma_B  \right) = 0 \;.
 \label{eq:constr-gam}
 \end{equation}
Since $G^{AB}$ is positive definite, we conclude that this can only be fulfilled for arbitrary values of $ \tilde{q}_A$ if $\tilde{\gamma}_B = \gamma_B$.
On the other hand, varying the first line of ${\cal L}_0$ with respect to $X^A$ and 
using \eqref{solXA}
gives
\begin{equation}
- 2 \mu \,  r \, e^{U_0} \, \tilde{q}_A  \left( 2 \, \delta X^A + G^{AC} \, \delta G_{CD} \, X^D \right) \;,
\end{equation}
which vanishes by virtue of $\delta G_{CD} \, X^D = 3 \, \delta X_C = - 2 \, G_{CD} \, \delta X^D$.
Thus, we conclude that the set of
variational equations derived from ${\cal L}_0$
is consistent.

Now we turn to ${\cal L}_2$, 
\begin{eqnarray}
{\cal L}_2&=&e^{2B+W + U_0-V_0+2 U_2}\bigg[- \frac12 \bigg(j \, e^{-2B-2W+V_0-U_2}+(W'-(U_0'+U_2'))\bigg)^2\nonumber\\
&&\qquad\qquad-  \left(B' - \frac12(U_0'+U_2' + W') + \frac32 X_A p^A \, e^{-2B+V_0-U_2} \right)^2 \nonumber\\
&& \qquad \qquad 
 + \frac13 \left[ 3 \left(B' + \frac12(U_0'+U_2' + W') \right) -  2 X^A h_A e^{V_0-U_2} + \frac32 X_A \, p^A \,
 e^{-2 B+V_0-U_2} \right]^2 \nonumber\\
&& \hskip -8mm  - G_{AB} \left( X'^A - \left[ \frac23 X^C (h_C e^{V_0-U_2} + G_{CD} p^D e^{-2B+V_0-U_2}) X^A - G^{AC} 
(h_C e^{V_0-U_2} +  G_{CD} p^D e^{-2B+V_0-U_2})
 \right] \right) \nonumber\\
 &&  \hskip -5mm \left( X'^B - \left[ \frac23 X^E (h_E e^{V_0-U_2} + G_{EF} p^F e^{-2B+V_0-U_2}) X^B - G^{BE} 
(h_E e^{V_0-U_2} +  G_{EF} p^F e^{-2B+V_0-U_2})
 \right] \right)
\bigg]
\nonumber\\
&& + 2 \left( e^{2B+W + U_0+U_2} \,\left(  X^A h_A - \frac32 X_A p^A \, e^{-2B} \right)
\right)'  \nonumber\\
&&
-\left(j \, e^{-W+U_0+U_2}\right)'+2 e^{W + U_0+V_0} \, h_A p^A \;.
\end{eqnarray}
This yields the first-order flow equations 
\begin{eqnarray}\label{eomdyonicsplit2}
(X^A)'&=&\frac23 X^C (h_C e^{V_0-U_2} + G_{CD} p^D e^{-2B+V_0-U_2}) X^A - G^{AC} 
(h_C e^{V_0-U_2} +  G_{CD} p^D e^{-2B+V_0-U_2})\,,\nonumber\\
  W'&=&U_0'+U_2'- j \, e^{-2B-2W+V_0-U_2}\,,\nonumber\\
  0&=&B' - \frac12(U_0'+U_2' + W') + \frac32 X_A p^A \, e^{-2B+V_0-U_2}\,,\nonumber\\
  0&=&3 \left(B' + \tfrac12(U_0'+U_2' + W') \right) -  2 X^A h_A e^{V_0-U_2} + \frac32 X_A \, p^A \,
 e^{-2 B+V_0-U_2}\,,
\end{eqnarray}
as well as the constraint
\begin{equation}
h_Ap^A=0\,.
\label{hoconstr}
\end{equation}
We have thus derived two sets of first-order flow equations (one derived from ${\cal L}_0$ and the other derived
from ${\cal L}_2$) that need to be mutually consistent.  Consistency of these two sets implies certain relations
which we now derive.

Adding the third and fourth equation of \eqref{eomdyonicsplit2} gives
\begin{eqnarray}
 B'  =  \frac13 X^A h_A e^{V_0-U_2} -  X_A \, p^A \,  e^{-2 B+V_0-U_2}\,.
\end{eqnarray}
Combining this with the first equation of \eqref{eq:eqs-L0}
yields
\begin{eqnarray}
 U_0'  =  - \frac23 X^A h_A e^{V_0-U_2} + 2  X_A \, p^A \,  e^{-2 B+V_0-U_2}\,.
 \label{eq:u0-p-h}
\end{eqnarray}
Comparing with the second equation of \eqref{eq:eqs-L0} yields
\begin{eqnarray}
 \frac13 e^{-2B-W} X^A \tilde{q}_A    =  - \frac13 X^A h_A e^{-U_2} +  X_A \, p^A \,  e^{-2 B-U_2}\,.
 \label{eq:xq-rel}
\end{eqnarray}

Contracting the first equation of \eqref{eomdyonicsplit2} with $G_{AB}$ results in 
\begin{eqnarray}
X_A'= -  (\frac23 X^C h_C e^{V_0-U_2} + X_C p^C e^{-2B+V_0-U_2}) X_A + \frac23 
(h_A e^{V_0-U_2} +  G_{AD} p^D e^{-2B+V_0-U_2})\,, \nonumber\\
\end{eqnarray}
Using \eqref{eq:u0-p-h} gives
\begin{eqnarray}
X_A'= \left(U_0' - 3 X_C p^C e^{-2B+V_0-U_2}\right) X_A  + \frac23 
(h_A e^{V_0-U_2} +  G_{AD} p^D e^{-2B+V_0-U_2})\,.
\label{eq:X'-anot}
\end{eqnarray}

Using $W' = - \tfrac12 U_0'$ (from the first equation of \eqref{eq:eqs-L0})
in the second equation of \eqref{eomdyonicsplit2} gives
\begin{equation}
U_2'=-\frac32 \, U_0'+ j \, e^{-2B-2W+V_0-U_2}\,,
\label{u2u0}
\end{equation}
which expresses $U_2$ in terms of $U_0$.

The second equation of \eqref{eomdyonicsplit2} can be rewritten as 
\begin{equation}
U_0'+U_2'+ W' = 2 W' + j \, e^{-2B-2W+V_0-U_2}
\end{equation}
which, when inserted into the third equation of \eqref{eomdyonicsplit2}, gives
\begin{equation}
- j \, e^{-2W}+3  X_Ap^A =0\,.
\end{equation}
Using $- 2 W = U_0$ (ignoring an additive constant) as well as \eqref{solXA}, this results in
\begin{equation}
j
= H_A \, p^A \;.
\end{equation}
Since the left hand side is constant, we conclude that
\begin{equation}
\tilde{q}_A p^A = 0 \;\;\;,\;\;\; j = \gamma_A  p^A \;.
\label{constraintsqpj}
\end{equation}

Finally, comparing the first equation of \eqref{eomdyonicsplit0} with \eqref{eq:X'-anot}
\begin{equation}
- 3 X_C p^C e^{-2B-U_2}  X_A  + \frac23 
(h_A e^{-U_2} +  G_{AD} p^D e^{-2B-U_2}) + \frac23 \tilde{q}_A e^{-2B- W }  = 0 \;.
\label{eq:constr-X}
\end{equation}
Note that
the contraction of \eqref{eq:constr-X} with $X^A$ gives back \eqref{eq:xq-rel}. On the other hand, 
contracting \eqref{eq:constr-X} with $p^A$ and using \eqref{hoconstr}  and \eqref{constraintsqpj} gives
\begin{equation}
3  \left( X_A p^A \right)^2  = \frac23 p^A G_{AB} \, p^B \;.
\label{eq:contr-rel}
\end{equation}
Observe that both sides are positive definite. 
This relation is, for instance, satisfied for the STU-model $X^1 X^2 X^3 = 1 $.
Inserting the relation
\begin{equation}
G_{AB} = - \frac12 C_{ABC} X^C + \frac92 X_A X_B 
\end{equation}
into \eqref{eq:contr-rel} we obtain
\begin{equation}
X^A C_{ABC} p^B p^C = 0 \;.
\end{equation}

Thus, we conclude that the two sets \eqref{eq:eqs-L0} and \eqref{eomdyonicsplit2}
are mutually consistent, provided that \eqref{eq:constr-X} is satisfied and the constraints
\eqref{hoconstr} and \eqref{constraintsqpj} hold.

In the following, we solve the first-order flow equations for the case when $p^A=0$.  Then $j=0$,
so that from \eqref{u2u0} we obtain $U_2' = -\frac32 \, U_0'$, which also equals $W' + 2 B'$ by virtue of 
the first equation of 
\eqref{eq:eqs-L0}.  Thus $U_2 = W + 2 B$, up to an additive constant.  Then, \eqref{eq:constr-X}
is satisfied provided we set $\tilde{q}_A = - h_A$.  Summarizing, when $p^A =0$, we obtain
\begin{eqnarray} \label{summarizesol}
\frac14 G^{AB} {q}_A {q}_B &=& \mu \, G^{AB} \, h_A\gamma_B \;, \nonumber\\
e^{-U_0} X_A  &=&    \frac13 \, H_A \;\;\;,\;\;   H_A = \gamma_A + r^2 \, h_A\;, \nonumber\\
e^{-U_0} &=& \frac13 \, H_A  X^A \;,  \nonumber\\
B &=&  W = - \frac12 U_0 \;, \nonumber\\
U_2 &=& - \frac32 U_0 \;, \nonumber\\
V_0 &=& - \frac12 U_0 + \log r \;, \nonumber\\
f &=& - \mu \, r^2 + g^2 e^{- 3 U_0} \;, \nonumber\\
(e^A)' &=& \frac12 \, e^{2U_0} \,r \, G^{AB} q_B \;.
\label{eq:sol-bcs}
\end{eqnarray}
This is the black brane analog of the black hole solutions discussed in \cite{Cardoso:2008gm}.

For this class of solutions, 
we now check the Hamiltonian constraint
\begin{equation}
R_{tt} + \frac{\delta {\cal L}_{M}}{\delta g^{tt}} - \frac12 g_{tt} \left(R + {\cal L}_{M} \right) = 0 \;,
\label{eq:hamilt}
\end{equation}
where ${\cal L}_{M}$ denotes the matter Lagrangian.
Using
\begin{equation}
\sqrt{-g} \, g^{tt} \, \left(R_{tt} - \frac12 g_{tt} R \right) = - 3 \, e^{3B + U - V} \left(B'^2 + B' \, U' \right)
+ 3 \left( e^{3B+U-V} \, B' \right)' \;,
\end{equation}
as well as
\begin{eqnarray}
\frac{\delta {\cal L}_{M}}{\delta g^{tt}} - \frac12 g_{tt} \, {\cal L}_{M} = \frac12 \, g_{tt} \, 
\left[ e^{-6B} \, G^{AB} q_A q_B + e^{-2V} \, G_{AB} X'^A X'^B + g^2  \left(G^{AB} - 2 X^A X^B \right) h_A h_B \right]\;,
\nonumber\\
\end{eqnarray}
where we replaced the electric fields by their charges, 
we obtain for \eqref{eq:hamilt}, 
\begin{equation}
{\cal L}_0 + {\cal L}_2 - 6\left( e^{3B+U-V} \, B' \right)' = 0 \;.
\end{equation}
Imposing the first-order flow equations, this reduces to 
\begin{eqnarray}
&& \left(  e^{U_0} f_0 \, \tilde{q}_A \, X^A + 3 \mu \, B \right)' 
+  g^2 \left( e^{3B+U_0+U_2} \, \tilde{q}_A \, X^A \right)' 
+ 3 \left( e^{3B+U_0-V_0} \, f \, \, B' \right)' \nonumber\\
&&
+ \mu \,  r \, e^{U_0} \, \tilde{q}_A \, X^A 
 = 0 \;.
 \label{eq:hamilt2}
 \end{eqnarray}
Then, using the second equation of \eqref{eq:eqs-L0} and
\begin{eqnarray}
3 \, r \, B'' = - U'_0 \, e^{U_0} \, r^2 \, \tilde{q}_A \, X^A - e^{U_0} \, r \, \tilde{q}_A \, X^A
- e^{U_0} \, r^2 \, \tilde{q}_A \, X'^A \;,
\end{eqnarray}
we find that \eqref{eq:hamilt2} is satisfied on a solution to the first-order flow equations.
Thus, the Hamiltonian constraint  \eqref{eq:hamilt} does not lead to any further constraint.

The class of static solutions \eqref{eq:sol-bcs} was obtained long time ago in \cite{Behrndt:1998jd} by solving the equations of motion. Their mass density is determined by $\mu$.
Let us consider a black solution, with the horizon located at $f(r_H) =0$ with $e^{2 U_0}(r_H) \neq 0$. 
 Its temperature is then given by
\begin{equation}
T_H = \left[\frac{e^{U_0 - V_0}}{4 \pi} |f'| \right]_{r = r_H}
= \left[\frac{e^{\tfrac32 U_0 }}{4 \pi \, r} |f'| \right]_{r = r_H} \;,
\label{eq:temp}
\end{equation}
where we used \eqref{summarizesol} to express $V_0$ in terms of $U_0$.
The horizon condition $f(r_H)=0$ gives
\begin{equation}
\mu \, r^2_H = g^2 \,e^{- 3 U_0(r_H)} \;,
\label{eq:hor}
\end{equation}
and hence $r_H \neq0$.  For the solution to be extremal, its temperature has to vanish.
Imposing
$f'(r_H) =0$ results in
\begin{equation}
\mu \, r_H + \frac32 g^2 \, e^{-3U_0(r_H)} \, U_0'(r_H) = 0 \;.
\label{eq:T=0}
\end{equation}
Combining both equations gives
\begin{equation}
1 + \frac32 \, r_H \, U_0'(r_H) = 0 \;.
\label{eq:cond-extr}
\end{equation}
As an example, consider the STU-model $X^1 X^2 X^3 =1$ and take $\gamma_A = c \, h_A$
with $c>0$, so that
$H_A = h_A \, H$ with $H= c + r^2$.  Then \begin{eqnarray} \label{xh_sols}
&& e^{-3 U_0} = H_1 H_2 H_3 = h_1 h_2 h_3 \, H^3 \quad , \quad X^1 = \frac{(H_2 H_3)^{1/3}}{H_1^{2/3}} = 
 \frac{(h_2 h_3)^{1/3}}{h_1^{2/3}} \;, \nonumber\\
 &&
X^2 = \frac{(H_1 H_3)^{1/3}}{H_2^{2/3}}=
 \frac{(h_1 h_3)^{1/3}}{h_2^{2/3}}
 \quad , \quad X^3 = \frac{(H_1 H_2)^{1/3}}{H_3^{2/3}} = \frac{(h_1 h_2)^{1/3}}{h_3^{2/3}}\ .
\end{eqnarray}
The scalars $X^A$ are thus constant.
We set $g=1$ and  $\alpha^3 = h_1 h_2 h_3  >0$.
The horizon condition \eqref{eq:hor} yields
\begin{equation}
\mu \, H(r_H) - \alpha^3 H^3(r_H) = \mu \, c \;,
\label{hor-stu-cond}
\end{equation} 
while 
the condition \eqref{eq:T=0} gives
\begin{equation}
r_H \left(\mu - 3 \alpha^3 \, H^2(r_H) \right) = 0 \;,
\end{equation}
which implies (we take $r_H \neq 0$),
\begin{equation}
\mu =  3 \alpha^3 \, H^2(r_H)  \;.
\label{eq:value-mu}
\end{equation} 
Inserting \eqref{eq:value-mu} into \eqref{hor-stu-cond}
 gives
\begin{equation}
2 \alpha^3 H^3(r_H) = \mu \, c \;.
\label{eq:rel1}
\end{equation}
Combining \eqref{eq:rel1} with the first equation of \eqref{eq:sol-bcs}, which takes the form
\begin{equation}
\mu \, c =  \frac14 \frac{q_A \, G^{AB} \, q_B }{h_A \, G^{AB} \, h_B} \;,
\end{equation}
yields the entropy density as 
\begin{equation}
{\cal S} = \frac14 e^{3B(r_H)} = \frac14 \left( \alpha^3 H^3 (r_H) \right)^{1/2} 
= \frac14 \left( \frac{\mu \, c}{2}  \right)^{1/2}
= \frac18 \, 
\left( \frac12 \, \frac{q_A \, G^{AB} \, q_B }{h_A \, G^{AB} \, h_B}  \right)^{1/2} \;.
\end{equation} 
Extremal electric solutions of this type have been considered recently in \cite{Donos:2011ff,Donos:2011pn}.


\end{appendix}

\bibliographystyle{JHEP}
\bibliography{references}

\end{document}